\documentclass[aps,prd,reprint,notitlepage,groupedaddress,longbibliography,nofootinbib]{revtex4-1}
\usepackage[utf8]{inputenc}
\usepackage{amsmath,amssymb,amsfonts}
\usepackage{graphicx}
\usepackage{color}
\usepackage[colorlinks=true,
linkcolor=blue,
urlcolor=black,
citecolor=blue]{hyperref}
\usepackage{tikz}
\usetikzlibrary{decorations.markings}
\usepackage{slashed}

\usepackage{tgtermes}

\def\be{\begin{equation}}
\def\ee{\end{equation}}
\def\bes{\begin{equation*}}
\def\ees{\end{equation*}}
\def\bea{\begin{eqnarray}}
\def\eea{\end{eqnarray}}

\def\f{\frac}

\def\mc{\mathcal}

\def\v[#1]{\boldsymbol{#1}}
\def\w[#1]{\widehat{#1}}
\def\vs[#1,#2]{\boldsymbol{{#1}_{#2}}}
\def\mes[#1]{d^{3}{#1}}
\def\del{\partial}
\def\<{\langle}
\def\>{\rangle}
\def\vec[#1]{\boldsymbol{#1}}
\def\vecs[#1,#2]{\boldsymbol{{#1}_{#2}}}

\newcommand{\B}[1]{{\bar{1}}}
\newcommand{\BD}[1]{{\dot \bar{1}}}
\newcommand{\half}{\frac{1}{2}}

\def\a{\alpha}
\def\b{\beta}
\def\d{\delta}
\def\D{\Delta}
\def\e{\epsilon}

\def\g{\gamma}
\def\G{\Gamma}

\def\l{\lambda}
\def\L{\Lambda}
\def\m{\mu}
\def\n{\nu}
\def\N{\nabla}
\def\o{\omega}
\def\O{\Omega}
\def\p{\phi}

\def\s{\sigma}

\def\t{\tau}
\def\th{\theta}

\def\coeff#1#2{{\textstyle {\frac {#1}{#2}}}}

\begin{document}

\title{Equilibrium thermodynamic susceptibilities for a dense degenerate Dirac field}

\author{Ashish Shukla}
\email{ashish@uvic.ca}
\affiliation{Department of Physics \& Astronomy, University of Victoria, Victoria, BC, Canada V8P 5C2}

\begin{abstract}
Parity preserving relativistic fluids in four spacetime dimensions admit seven independent thermodynamic susceptibilities at the second order in the hydrodynamic derivative expansion. We compute all parity even second order thermodynamic susceptibilities for a free massive Dirac field at zero temperature and a non-zero chemical potential, based on the Kubo formulas reported in \cite{Kovtun:2018dvd}. We also compute the second order constitutive relations for the energy-momentum tensor and the conserved current in the absence of external gauge fields.
\end{abstract}

\maketitle
\section{Introduction}
\label{Intro}
Hydrodynamics provides an effective low energy description of physical systems near thermal equilibrium \cite{landau2013fluid, Kovtun:2012rj, Romatschke:2017ejr}. The description is valid on length scales much larger than the typical microscopic length scales associated with the system, such as the mean free path or the correlation length. In the hydrodynamic regime, the conserved currents of the system admit a derivative expansion in terms of the hydrodynamic variables such as temperature, chemical potential, fluid velocity etc. The dynamical equations of hydrodynamics express the conservation of the energy-momentum tensor and other conserved currents in the system.

At each order in the hydrodynamic derivative expansion one can have several transport coefficients quantifying the response of the system to external perturbations. Some of these transport coefficients are dissipative in nature, contributing to entropy production out of equilibrium, whereas others are non-dissipative and exist even in the limit of thermal equilibrium \cite{Baier:2007ix, Bhattacharyya:2008jc, Jensen:2011xb}. Interestingly, as was first noted in \cite{Banerjee:2012iz, Jensen:2012jh}, the non-dissipative transport coefficients follow from an equilibrium generating functional which itself admits a derivative expansion, with various thermodynamic susceptibilities appearing as coefficients of different terms in the expansion. The variation of the equilibrium generating functional with respect to external sources gives the constitutive relations for the conserved currents in the system, which contain various transport coefficients that do not vanish in the equilibrium limit. Thus, in thermal equilibrium, one can think of the susceptibilities entering the generating functional to be the more fundamental objects, and the transport coefficients appearing in the constitutive relations to be derived entities.

For relativistic fluids with a conserved current in equilibrium in four spacetime dimensions there are no thermodynamic susceptibilities at the first order in derivatives, whereas at the second order there are nine \cite{Banerjee:2012iz}. Out of the nine susceptibilities, seven are parity even. Kubo formulas for all the second order thermodynamic susceptibilities in terms of the correlations functions of the conserved currents were derived in \cite{Kovtun:2018dvd}. As it turns out, the parity-even susceptibilities can be computed using zero-frequency flat space two-point correlation functions of the energy-momentum tensor and the conserved current in the absence of sources, whereas the parity-odd susceptibilities require three-point functions. 

In this paper we make use of the Kubo formulas of \cite{Kovtun:2018dvd} to compute the seven parity-even equilibrium thermodynamic susceptibilities for a free massive Dirac field at zero temperature and non-zero chemical potential. Such a dense and degenerate matter configuration is of interest in astrophysical settings, such as white dwarfs and neutron stars \cite{Shapiro:1983du, Glendenning:1997wn, Vuorinen:2018qzx, Annala:2019eax}, and in QCD \cite{ Alford:2007xm, Kurkela:2009gj, Ghisoiu:2016swa}. We also compute the constitutive relations for the energy-momentum tensor and the conserved current, when the theory is not coupled to an external gauge field conjugate to the current.

The paper is organized as follows. Following \cite{Kovtun:2018dvd}, sections \ref{gf} and \ref{constitutive} review the formalism of equilibrium generating functional, the derivative expansion, and quote the Kubo formulae for the parity-even second order thermodynamic susceptibilities. Section \ref{Diracsetup} then reviews the formalism for Dirac fields, mentioning in particular the energy-momentum tensor and the two-point function which we make use of in our computations. The thermodynamic susceptibilities are then computed in section \ref{thsusc}, which also includes a discussion of the conformal limit. We compute the second order constitutive relations and end with a discussion in section \ref{discussion}. Appendix \ref{FTE} contains a discussion of the effects of non-zero temperature.

\section{Basic Setup}
\label{setup}
\subsection{Thermal equilibrium and the generating functional}
\label{gf}
We consider a macroscopic system with a conserved current in equilibrium. The system can be coupled to an external non-dynamical metric $g_{\m\n}$, and an external non-dynamical gauge field $A_\m$ corresponding to the conserved current. Equilibrium implies the presence of a timelike Killing vector $V^\m$. Coordinates in which $V^\m = (1, \v[0])$ correspond to the matter at rest. In the thermodynamic frame \cite{Jensen:2012jh}, the fluid four-velocity $u^\m$, the temperature $T$, and the chemical potential $\m$ can be defined via the Killing vector $V^\m$ as
\be
\label{fluvar}
u^\m = \f{V^\m}{\sqrt{-V^2}}, \quad T = \f{T_0}{\sqrt{-V^2}}, \quad \m = \f{V^\m A_\m + \L_V}{\sqrt{-V^2}}.
\ee
Here $T_0$ is the equilibrium temperature in the matter rest frame, and $\L_V$ is a gauge function introduced to ensure the gauge independence of $\m$. 
The condition for the system to be in equilibrium translates to
\be
\label{eqcond}
\mathfrak{L}_V g_{\m\n} = 0\, , \quad \mathfrak{L}_V A_{\m} + \del_\m \L_V = 0\, ,
\ee
where $\mathfrak{L}_V$ denotes the Lie derivative with respect to $V^\m$.

As has been shown in \cite{Jensen:2012jh, Banerjee:2012iz}, physical properties of the fluid in equilibrium, in particular the zero-frequency correlation functions over length scales much bigger than the length scales associated with the microscopic dynamics of the fluid, are encoded in a generating functional $\mathcal{W}$. The generating functional is a functional of the external sources. For the system under consideration, the generating functional is a functional of the background metric $g_{\m\n}$ and the background gauge field $A_\m$, which act as sources for the energy-momentum tensor and the conserved current, respectively. The generating functional can be expressed as
\be
\label{defgfunc}
\mc{W}[g_{\m\n}, A_\m] = \int d^4x \, \sqrt{-g} \, \mc{F}[g_{\m\n}, A_\m],
\ee
where $\mc{F}[g_{\m\n}, A_\m]$ is a local function of the sources. The variation of the generating functional gives
\be
\begin{split}
\label{vargf}
&\d\mc{W}[g_{\m\n}, A_{\m}] = \\
&\quad\half \int d^4x \sqrt{-g}\, T^{\m\n} \d g_{\m\n} + \int d^4x \sqrt{-g}\, J^\m \d A_\m ,
\end{split}
\ee
with $T^{\m\n}, J^\m$ respectively denoting the energy-momentum tensor and the conserved current.

We assume that the system is free of any quantum anomalies, in which case the generating functional is both diffeomorphism as well as gauge invariant. This translates to
\begin{align}
\N_\m T^{\m\n} &= F^{\n\l} J_{\l}, \label{fleq1}\\
\N_\m J^\m &= 0, \label{fleq2}
\end{align}
where $F_{\m\n} = \del_\m A_\n - \del_\n A_\m$ is the field strength tensor for the background field $A_\m$. 

In the hydrodynamic regime, when the external sources vary over length scales much larger than the microscopic length scales associated with the fluid such as the mean free path, the density $\mc{F}[g_{\m\n}, A_\m]$ admits a derivative expansion. The expansion is in terms of the derivatives of the sources as well as the fluid variables defined in eq.\! \eqref{fluvar}. Thus, to compute the generating functional one is interested in finding out all the diffeomorphism and gauge invariant objects that can be constructed out of the sources and the fluid variables upto a given order in derivatives.

For instance, to the zeroth order in the derivatives we have only two invariants, $T$ and $\m$. The generating functional then takes the form
\be
\label{genfunc0}
\mc{W}[g_{\m\n}, A_\m] = \int d^4x \sqrt{-g}\, p(\m, T) + \cdots,
\ee
where $p(\m, T)$ is the equilibrium pressure of the system\footnote{It is straight forward to see that $p(\m,T)$ is the equilibrium pressure. The energy-momentum tensor and the current following from the variation of eq. \eqref{genfunc0} take the form
\begin{align*}
&T^{\m\n} = (\e + p) u^\m u^\n + p g^{\m\n} + \cdots ,\\
&J^\m = n u^\m + \cdots ,
\end{align*}
which have the form of a perfect fluid, with $\e = - p + T \del p/\del T + \m \,\del p/ \del \m$ being the equilibrium energy density, and $n = \del p/\del \m$ being the equilibrium charge density, provided $p(\m, T)$ is identified as the equilibrium pressure.}, and the dots denote terms with higher derivatives. 

There are no invariants at the first order in derivatives in equilibrium, as the equilibrium conditions eq.\! \eqref{eqcond} guarantee that the scalars $u^\l \del_\l T, u^\l \del_\l \m , \N.u$ all vanish. 

At the second order in derivatives, interestingly, there are nine invariants \cite{Banerjee:2012iz}. In terms of these the generating functional can be represented as
\be
\begin{split}
\mc{W}&[g_{\m\n}, A_\m] = \\
&\int d^4x \sqrt{-g}\left[ p(\m, T)+ \sum^9_{n = 1} f_n(\m, T) \, s_n^{(2)} + \cdots\right],
\end{split}
\ee
where $s_n^{(2)}$ denotes the $n^{th}$ second order invariant, and $f_n(\m, T)$ denotes the corresponding thermodynamic susceptibility \cite{Kovtun:2018dvd}. In terms of the acceleration $a^\m = u^\n \N_\n u^\m$, the vorticity $\O^\m = \e^{\m\n\a\b} u_\n \N_\a u_\b$,\footnote{We have $\e^{\m\n\a\b} = \varepsilon^{\m\n\a\b}/\sqrt{-g}$, with $\varepsilon^{0123} = 1$.} the electric field $E^\m = F^{\m\n} u_\n$, and the magnetic field $B^\m = \half \e^{\m\n\a\b} u_\n F_{\a\b}$, the generating functional written in terms of the nine invariants at the second order is\footnote{We count the metric and the gauge field as $\mc{O}(1)$ quantities. Consequently the Ricci scalar is $\mc{O}(\del^2)$, and the electric and magnetic fields are $\mc{O}(\del)$. The derivative counting may differ depending upon the nature of the fluid. For instance, if the fluid is insulating and has no free charges, than the electric field inside it is not screened and can be strong, in which case it is to be counted as $\mc{O}(1)$ rather than $\mc{O}(\del)$. Similarly, in magnetohydrodynamics the magnetic field can be strong and has to be counted as $\mc{O}(1)$. See \cite{Kovtun:2016lfw, Hernandez:2017mch} for additional discussion.}
\be
\begin{split}
\label{gfsecord}
\mc{W}[g_{\m\n}, A_\m] = \int& d^4x \sqrt{-g} \, \Big[ p(\m,T) + f_1 R + f_2 a^2 \\
&+ f_3 \O^2 + f_4 B^2 + f_5 B\cdot\O + f_6 E^2 \\
&\quad\,\,+ f_7 E\cdot a + f_8 E\cdot B + f_9 B\cdot a \Big],
\end{split}
\ee
where $R$ denotes the Ricci scalar of the background geometry. The susceptibilities $f_n(\m, T)$ are to be determined from the microscopic theory governing the system, just like the pressure. The seven susceptibilities $f_1, \cdots f_7$ are parity even, whereas $f_8, f_9$ are parity odd. Our primary focus in the discussion that follows will be on the parity-even susceptibilities.

\subsection{Constitutive relations and the Kubo formulae}
\label{constitutive}
In terms of the fluid four-velocity $u^\m$, the energy-momentum tensor for the fluid can be decomposed as
\be
\label{stressdecomp}
T^{\m\n} = \mc{E} u^\m u^\n + \mc{P} \D^{\m\n} +\mc{Q}^\m u^\n + \mc{Q}^\n u^\m + \t^{\m\n},
\ee
where $\D_{\m\n} = g_{\m\n} + u_\m u_\n$ is a projector that projects orthogonal to the fluid four-velocity $u^\m$. In the decomposition eq.\ \eqref{stressdecomp} for $T^{\m\n}$, the energy density $\mc{E} = u_\m u_\n T^{\m\n}$, the pressure $\mc{P} = \f{1}{3} \D_{\m\n} T^{\m\n}$, the energy flux $\mc{Q}^\m = - \D^{\m\a} T_{\a\b} u^\b$, and $\tau^{\m\n} = T^{\langle\m\n\rangle}$, where the angle brackets on a pair of indices denote the symmetric transverse-traceless projection,
\bes
A_{\langle\m\n\rangle} = \half \left(\D_{\m\a} \D_{\n\b} + \D_{\m\b} \D_{\n\a} - \f{2}{3} \D_{\m\n} \D_{\a\b} \right) A^{\a\b}.
\ees
Clearly, the energy flux is orthogonal to the fluid four-velocity, $u_\m \mc{Q}^\m = 0$, and $\t^{\m\n}$ is a symmetric tensor both orthogonal to the fluid four-velocity as well as traceless, $u_\m \t^{\m\n} = 0, g_{\m\n} \t^{\m\n} = 0$.

In the hydrodynamic regime, the quantities $\mc{E}, \mc{P}, \mc{Q}^\m$ and $\t^{\m\n}$ admit derivative expansions in terms of the derivatives of the sources and the fluid variables, called the constitutive relations. In the limit when the fluid is in equilibrium, we also have the generating functional eq.\ \eqref{defgfunc} at our disposal, whose variation with respect to the metric gives the energy-momentum tensor, eq.\ \eqref{vargf}. Consider for instance a fluid with a conserved charge in equilibrium, with the associated chemical potential being non-zero but the background gauge field set to zero. Upto second order in derivatives the generating functional of such a fluid then has the form of eq.\ \eqref{gfsecord}, with the background fields $E, B$ set to zero. Its variation with respect to the metric gives the constitutive relations \cite{Kovtun:2018dvd}
\begin{widetext}
\begin{subequations}
\label{constrel}
\begin{align}
&{\mc{E}}  =  \e + \left(f_1' - f_1\right) R + \left(4 f_1' + 2 f_1'' - f_2 - f_2'\right)a^2 + \left(f_1' - f_2 - 3f_3 + f_3'\right) \O^2 - 2\left(f_1 + f_1' - f_2\right) u^\a R_{\a\b} u^\b \,, \label{crele}\\
&{\mc{P}}  =  p + \f{1}{3} f_1 R - \f{1}{3} \left(2f_1' +f_3\right) \O^2 -\f{1}{3} \left(2f_1' + 4f_1'' - f_2\right) a^2 +\f{2}{3} \left(2f_1' - f_1\right) u^\a R_{\a\b} u^\b\,, \label{crelp}\\
&{\mc{Q}}_\m  = \left(f_1' + 2f_3'\right) \e_{\m\n\a\b}a^\n u^\a \O^\b + \left(2f_1 + 4f_3\right) \D_\m^{\,\n} R_{\n\s} u^\s\,, \label{crelq}\\
&{\mc{\t}}_{\m\n}  =  \left(4f_1' + 2f_1'' - 2f_2\right) a_{\langle\m} a_{\n\rangle} -\f{1}{2} \left(f_1' -4f_3\right) \O_{\langle\m} \O_{\n\rangle} + 2f_1'\, u^\a R_{\a\langle \m\n \rangle\b} u^\b -2f_1 R_{\langle \m\n\rangle}\, , \label{crelt}
\end{align}
\end{subequations}
\end{widetext}
where we have used the notation
\begin{align*}
f_n' &=  T f_{n,T} + \mu f_{n,\mu}\, ,\\
f_n''&= T^2 f_{n,T,T} + 2\mu T f_{n,T,\mu} + \mu^2 f_{n,\mu,\mu}\, ,
\end{align*}
with a comma denoting a partial derivative with respect to the argument following. The constitutive relations eq.\ \eqref{constrel} allow for a straight forward computation of the equilibrium correlation functions of the energy-momentum tensor. For instance, the equilibrium two-point function is defined as the variation of the equilibrium one-point function with respect to the source. For matter at rest on a flat background, we have
\be
\label{defvarT1}
\d_g\left( \sqrt{-g} \, T^{\m\n}\right) = \half \, G_{T^{\m\n} T^{\a\b}}(\o = 0, \v[k]) \, \d g_{\a\b}(\v[k]),
\ee
where we have decomposed the background metric as $g_{\m\n} = \eta_{\m\n} + \d g_{\m\n} (\v[k]) \, e^{i \v[k]\cdot\v[x]}$. Using eq.\ \eqref{defvarT1}, the constitutive relations eq.\ \eqref{constrel} imply Kubo formulae for the second order susceptibilities in terms of zero-frequency two-point correlation functions of the energy-momentum tensor, given by \cite{Kovtun:2018dvd}
\begin{align}
\label{Kubof1}
& f_1 = - \, \half \lim_{\v[k]\to0} \frac{\partial^2}{\partial k_z^2} G_{T^{xy} T^{xy}} ,\\
\label{Kubof2}
& f_2 = \f{1}{4} \lim_{\v[k]\to0} \frac{\partial^2}{\partial k_z^2} \left( G_{T^{tt} T^{tt}} + 2 G_{T^{tt} T^{xx}} - 4 G_{T^{xy} T^{xy}}\right) ,\\
\label{Kubof3}
& f_3 = \f{1}{4} \lim_{\v[k]\to0} \frac{\partial^2}{\partial k_z^2} \left( G_{T^{tx} T^{tx}} + G_{T^{xy} T^{xy}} \right).
\end{align}
The Kubo formulas above are not the only ones we obtain by varying the energy-momentum tensor; for instance, $f_1$ can also be computed by using $4f_1 = \lim_{\v[k]\rightarrow 0} \del^2 G_{T^{xx}T^{yy}}/\del k_z^2$. However, the formulas above are simple enough to be used for the computation of the susceptibilities as we shall see later. Note also that $G_{T^{\m\n} T^{\a\b}}$ may include contact term contributions; see \cite{Moore:2010bu, Moore:2012tc} for a discussion. 

Similar to eq.\ \eqref{stressdecomp}, the  $U(1)$ current can be expressed as
\be
\label{defJ}
J^\m = \mc{N} u^\m + \mc{J}^\m,
\ee 
where the charge density $\mc{N} = - u_\m J^\m$, and the spatial current $\mc{J}^\m = \D^\m_{\,\n} J^\n$, with $u_\m \mc{J}^\m = 0$. In equilibrium, in the spirit of eq.\ \eqref{defvarT1}, the two-point functions of the current are defined in terms of the variation of the one-point functions,
\begin{align}
\d_g\left( \sqrt{-g} \, J^{\m}\right) &= \half \, G_{J^{\m} T^{\a\b}}(\o = 0, \v[k]) \, \d g_{\a\b}(\v[k]), \label{defvarJ1}\\
\d_A\left( \sqrt{-g} \, J^{\m}\right) &= G_{J^{\m} J^{\n}}(\o = 0, \v[k]) \, \d A_{\n}(\v[k]), \label{defvarJ2}
\end{align}
where the variation in the background field is $\d A_\l(\v[x]) \equiv \d A_\l(\v[k]) e^{i\v[k]\cdot\v[x]}$. From the second order generating functional eq.\ \eqref{gfsecord} one can compute the current by varying it with respect to $A_\m$, eq.\ \eqref{vargf}, and then vary the current with respect to the metric and the gauge field, eqs.\ \eqref{defvarJ1}, \eqref{defvarJ2}, giving Kubo formulae for the susceptibilities $f_4, \cdots f_7$, which are (see \cite{Kovtun:2018dvd} for details)
\begin{align}
&f_4 = \f{1}{4} \lim_{\v[k]\to0} \frac{\partial^2}{\partial k_z^2} G_{J^{x} J^{x}}\,, \label{Kubof4}\\
&f_5 = \half \lim_{\v[k]\to0} \frac{\partial^2}{\partial k_z^2} G_{J^{x} T^{tx}}\,,\label{Kubof5}\\
&f_6 = \f{1}{4} \lim_{\v[k]\to0} \frac{\partial^2}{\partial k_z^2} G_{J^{t} J^{t}}\,,\label{Kubof6}\\
&f_7 = -\half \lim_{\v[k]\to0} \frac{\partial^2}{\partial k_z^2} (G_{J^{t} T^{tt}} + G_{J^{t} T^{xx}})\,\label{Kubof7}.
\end{align}
The constitutive relations for a fluid with a conserved current $J^\m$ not coupled to the corresponding background field are
\begin{subequations}
\label{constrelJ}
\begin{align}
&{\mc{N}} = n + f_{1,\m} R + \left(f_{2,\m} + f_7 + f_7'\right) a^2 \nonumber\\
&\quad\quad+ \left(f_{3,\m} - f_5 + \coeff12 f_7 \right) \O^2 - f_7\, u^\a R_{\a\b}u^\b ,\\
&{\mc{J}}^\m = -\left(f_5 + f_5'\right)\e^{\m\n\rho\s}u_\n a_\rho \O_\s +2f_5\D^{\m\n}R_{\n\l} u^\l,
\end{align}
\end{subequations}
where $n=\del p/\del\m$ is the charge density at zeroth-order.

\subsection{Review of the Dirac formalism}
\label{Diracsetup}
The focus of the present article is on computing the parity-even second order thermodynamic susceptibilities $f_1, \cdots f_7$ for a free massive  Dirac field at zero temperature and a non-zero chemical potential. In the present subsection we briefly review the formalism for Dirac fields and present the essential formulas we will need for computing the susceptibilities from two-point functions of the energy-momentum tensor and current, sec.\ \ref{constitutive}. The action for the free massive Dirac field on a curved spacetime background is given by
\be
\label{Dirac1}
S = \int d^4x \sqrt{-g} \, \bar{\Psi} \left( i {\G}^\m \N_\m - m\right) \Psi,
\ee
where ${\G}^\m(x)$ are the spacetime dependent $\g$-matrices defined via ${\G}^\m(x) = e^\m_a(x) \g^a$, with $\g^a$ being the spacetime independent $\g$-matrices and $e^\m_a(x)$ being the vierbein fields. The Clifford algebras satisfied by the $\g$-matrices are
\begin{align}
&\left\{ \g^a, \g^b\right\}= - 2 \eta^{ab} ,\\
&\left\{ {\G}^\m(x), {\G}^\n(x)\right\}= - 2 g^{\m\n}(x).
\end{align}
The covariant derivative of the spinor field appearing in the action eq.\eqref{Dirac1} is given by
\be
\label{derspin}
\N_\m \Psi = \del_\m \Psi + \half\, \o_{\m}^{\,ab} \s_{ab} \Psi,
\ee
with $\s_{ab} = \f{1}{4} \left[\g_a, \g_b\right]$, and the spin connection $\o_{\m}^{\,ab}$ given by
\be
\begin{split}
\label{spincon}
\o_{\m}^{\,ab} = \half \, e^{a\n} \left(\del_\m e^b_\n - \del_\n e^b_\m\right) - \half\, e^{b\n} \left(\del_\m e^a_\n - \del_\n e^a_\m\right) \\+ \half \, e^{a\n} e^{b\rho} \left(\del_\rho e^c_\n - \del_\n e^c_\rho\right) e_{c\m}.
\end{split}
\ee

For computing the susceptibilities we will need the energy-momentum tensor and the current following from the action eq.\eqref{Dirac1}. The energy-momentum tensor for the Dirac field can be computed by using $T^{\m\n} = \f{e^\n_a}{\sqrt{-g}} \f{\delta S}{\delta e_{\m a}}$, which gives \cite{dewitt2003global}
\be
\label{DiracStress1}
T^{\m\n} = \f{i}{4} \left( \N^\m \bar{\Psi} {\G}^\n \Psi - \bar{\Psi}{\G}^\n \N^\m \Psi \right) + \m \leftrightarrow \n.  
\ee
In flat space, the covariant derivatives in the above equations can be replaced by ordinary derivatives, and the spacetime dependent $\g$-matrices become spacetime independent. 

We will find it convenient to work in Euclidean signature rather then Lorentzian. The Euclidean Dirac action in flat space is given by
\be
\label{EDirac}
S_E = \int d^4x\,\bar{\Psi} (\tilde{\g}^\m \del_\m + m) \Psi,
\ee
where $\tilde{\g}^\m$ are the Euclidean gamma matrices such that $\tilde{\g}^0 = \g^0, \tilde{\g}^i = -i \g^i$. They satisfy the Clifford algebra
\be
\label{EClifford}
\{ \tilde{\g}^\m, \tilde{\g}^\n\} = 2 \d^{\m\n}. 
\ee
The trace identities of interest are
\begin{align}
&\text{Tr}\left[ \tilde{\g}^\m \tilde{\g}^\n\right] = 4 \d^{\m\n}\, ,\\
&\text{Tr}\left[ \tilde{\g}^\m \tilde{\g}^\n \tilde{\g}^\rho \tilde{\g}^\s\right] = 4 \left(\d^{\m\n} \d^{\rho\s} - \d^{\m\rho} \d^{\n\s} + \d^{\m\s} \d^{\n\rho}\right).
\end{align}
The Euclidean energy-momentum tensor has the form
\be
\label{EucStress}
T^{\m\n}_E = \f{1}{4} \left( \bar{\Psi} \tilde{\g}^\n \del^\m \Psi - \del^\m \bar{\Psi} \tilde{\g}^\n \Psi\right) + \m \leftrightarrow \n.
\ee

The conserved current for the global $U(1)$ symmetry of the Dirac action eq.\ \eqref{Dirac1} is $J^\m = \bar{\Psi} \g^\m \Psi$, which when written in Euclidean signature implies $J^0_E = \bar{\Psi} \tilde{\g}^{0} \Psi, J^i_E = i \bar{\Psi} \tilde{\g}^i \Psi$.
 
Finally, the Euclidean space two-point function has the form
\be
\label{2ptEdirac}
\langle \Psi_{\a}(k) \bar{\Psi}_{\b}(k')\rangle = (2\pi)^4 \delta^4(k + k') \f{\left(-i\slashed{k} + m\right)_{\a\b}}{k^2 + m^2}.
\ee

Notice that the above formulas were written for a Dirac field with a vanishing chemical potential. At a non-zero chemical potential $\m$ for the global $U(1)$ symmetry of the Dirac action, the Euclidean action becomes
\be
\label{EucDirac}
S_{E,\, \m \neq 0} = \int d^4x\,\bar{\Psi} (\tilde{\g}^\m \del_\m + m - \m \tilde{\g}_0) \Psi.
\ee 
The effect of the non-zero chemical potential can easily be accommodated into our computations by shifting the zeroth component of the Euclidean four-vector $k^\m$ in the two-point function eq.\ \eqref{2ptEdirac} to $k^0 \rightarrow k^0 + i \m$.

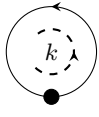
\begin{figure}[ht]
\begin{tikzpicture}[>=stealth]
\draw (0,0) [decoration={markings,mark=at position 1/4 with {\arrow[line width=0.2mm]{>}}},postaction={decorate}] circle (6mm);
\draw [dashed,thick,decoration={markings,mark=at position 1/2 with {\arrow[line width=0.2mm]{>}}},postaction={decorate}] ([shift=(-135:3mm)]0,0) arc (-135:135:3mm);
\draw [fill] (0,- 6mm) circle [radius=0.1];
\node at (0,0) {$k$};
\end{tikzpicture}
\caption{The one-loop diagram contributing to the zeroth-order energy density, pressure and charge density. The node corresponds to an insertion of the energy-momentum  tensor or the current.}
\label{onepoint}
\end{figure}

As a warm-up exercise, we can compute the zeroth-order equilibrium energy density, pressure and charge density for the free massive Dirac field at $T=0, \m \neq 0$ using the above results. The energy density is given by the one-point function $\e = \langle T^{00}(\o = \v[k] = 0)\rangle' = -\, \langle T^{00}_E(\o = \v[k] = 0)\rangle'$.\footnote{Components of the energy-momentum tensor in Lorentzian and Euclidean signatures are related via $T^{00} \leftrightarrow - T^{00}_E,\, T^{0i} \leftrightarrow - i T^{0i}_E,\, T^{ij} \leftrightarrow T^{ij}_E$.}\textsuperscript{,}\footnote{A $'$ denotes the removal of the factor  of $(2\pi)^4 \d^4(0)$ from the correlator.}
This can be thought of as evaluating the one-loop diagram of figure \ref{onepoint}. Using the expression for $T^{\m\n}_E$ from eq.\ \eqref{EucStress} and the two-point function eq.\eqref{2ptEdirac}, one finds that $\e = 0$ for $|\m| < m$, whereas for $|\m| > m$ we have
\begin{align}
\label{energy}
\e = \f{\m^4}{8\pi^2} \Bigg[ \bigg(2 &- \f{m^2}{\m^2}\bigg) \sqrt{1 - \f{m^2}{\m^2} } \nonumber \\
&- \f{m^4}{\m^4} \log\left(\f{|\m|+\sqrt{\m^2 - m^2}}{m}\right)\Bigg].
\end{align}
Similarly, the equilibrium pressure can be computed by using $p = \d_{ij}\langle T^{ij}(\o = \v[k] = 0)\rangle'/3$, which shows that for $|\m| > m$
\begin{align}
\label{pressure}
p = \f{\m^4}{24\pi^2} \Bigg[\bigg(2 &- \f{5m^2}{\m^2}\bigg) \sqrt{1 - \f{m^2}{\m^2} } \nonumber\\&+ \f{3m^4}{\m^4} \log\left(\f{|\m|+\sqrt{\m^2 - m^2}}{m}\right)\Bigg],
\end{align}
whereas for $|\m| < m$ we have $p = 0$.
\begin{figure}[ht]
\includegraphics[scale=0.90]{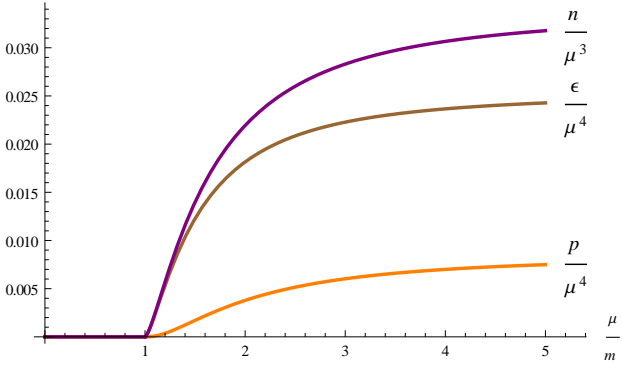}
\caption{The zeroth-order equilibrium energy density, pressure and charge density as a function of $\m/m$ at $T = 0$ for free massive Dirac fermions. Plots based on eqs.\ \eqref{energy}, \eqref{pressure} and \eqref{numbdensity}. $\e$ and $p$ are even functions of $\m$, whereas $n$ is odd.}
\label{zeroth}
\end{figure}

The equilibrium charge density is given by the formula $n = \langle J^0(\o = \v[k] = 0)\rangle'$, which implies that for $|\m| >m$
\be
\label{numbdensity}
n = \f{\text{Sign}(\m)}{3 \pi^2} \left(\m^2 - m^2 \right)^{3/2},
\ee
and $n = 0$ for $|\m| < m$. Clearly, $\e, p$ are even functions of $\m$, whereas $n$ is odd. Fig. \ref{zeroth} shows plots of $\e, p$ and $n$ for $\m>0$.

\section{Thermodynamic Susceptibilities}
\label{thsusc}
With the aid of the Kubo formulas discussed in sec. \ref{constitutive} we can now compute the equilibrium thermodynamic susceptibilities $f_1, \cdots f_7$ for the free massive Dirac field at zero temperature and non-zero chemical potential. The particle number density vanishes for $|\m| < m$ at $T = 0$, as can be seen from the Fermi distribution. Thus for $|\m|<m$ the system is essentially in its vacuum state. As will become clear from the computations below, all the susceptibilities vanish for $|\m| < m$.

\begin{figure}[t]
\begin{tikzpicture}[>=stealth]
\draw (0,0) [decoration={markings,mark=at position 1/4 with {\arrow[line width=0.2mm]{<}},mark=at position 3/4 with {\arrow[line width=0.2mm]{<}}},postaction={decorate}] circle (6mm);
\draw [fill] (6mm,0) circle [radius=0.1];
\draw [fill] (-6mm,0) circle [radius=0.1];
\draw [->] (-15mm,0) -- (-9mm,0);
\draw [->] (9mm,0) -- (15mm,0);
\draw [->] (-3mm,-8mm) -- (3mm,-8mm);
\draw [->] (-3mm,8mm) -- (3mm,8mm);
\draw [dashed,thick,decoration={markings,mark=at position 1/2 with {\arrow[line width=0.2mm]{>}}},postaction={decorate}] ([shift=(-135:3mm)]0,0) arc (-135:135:3mm);
\node at (-12mm,-3.0mm) {$k$};
\node at (12mm,-3.0mm) {$k$};
\node at (0,0) {$q$};
\node at (0,-11mm) {$q$};
\node at (0,11mm) {$k-q$};
\end{tikzpicture}
\caption{The one-loop diagram contributing to the susceptibilities. The nodes correspond to stress tensor and current insertions, depending upon the susceptibility being computed. $k$ denotes the external momentum, while $q$ is the momentum running in the loop.}
\label{twopoint}
\end{figure}
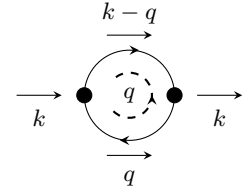

The Kubo formulas indicate that the susceptibilities can be computed from one-loop diagrams of the form shown in figure \ref{twopoint}. Consider for instance the computation of $f_1$, given by the Kubo formula eq.\ \eqref{Kubof1}. We need to first compute the quantity $G_{T^{xy} T^{xy}}$, same as $G_{T^{xy}_E T^{xy}_E}$, given in terms of the flat space zero-frequency two-point function via
\be
\langle T^{xy}(\o = 0,\v[k])\, T^{xy}(\o = 0,- \v[k])\rangle = (2\pi)^4 \d^4(0)\, G_{T^{xy} T^{xy}}.
\ee
From the expression for the energy-momentum tensor eq.\ \eqref{EucStress} it is straight forward to compute $T^{xy}_E(\o = 0,\v[k])$, which turns out to be
\be
\begin{split}
&T^{xy}_E(\o = 0,\v[k]) = \\
&- \f{i}{4} \int \f{d^4p}{(2\pi)^4} \Big[\left(\v[k]-2\,\v[p]\right)_x \bar{\Psi}(-p^0,\v[k]-\v[p])\, \tilde{\g}^y \, \Psi(p)\\
&\qquad\qquad\quad+\left(\v[k]-2\,\v[p]\right)_y \bar{\Psi}(-p^0,\v[k]-\v[p])\, \tilde{\g}^x \, \Psi(p)\Big],
\end{split}
\ee
where $p^\m$ is a Euclidean four-vector. Making use of eq.\ \eqref{2ptEdirac} we can now compute $G_{T^{xy}_E T^{xy}_E}$, which turns out to be
\begin{widetext}
%\begin{small}
\bes
G_{T^{xy}_E T^{xy}_E} = \int \f{d^4 q}{(2\pi)^4} \Bigg[\f{q_x^2 + q_y^2}{\big((q^0 + i\m)^2 + \v[q]^2 + m^2 + k_z^2 - 2 q_z k_z\big)} - \f{q_z k_z (q_x^2 + q_y^2) + 8 q_x^2 q_y^2}{\big((q^{0} + i\m)^2 + \v[q]^2 + m^2\big)\big((q^0 + i\m)^2 + \v[q]^2 + m^2 + k_z^2 - 2 q_z k_z\big)}\Bigg],
\ees
%\end{small}
\end{widetext}
where, without any loss of generality, we have assumed that $\v[k] = (0,0,k_z)$. Notice that we have shifted $q^0 \rightarrow q^0 + i\m$ to take into account the non-zero value of the chemical potential $\m$. To compute $f_1$ we now act $G_{T^{xy}_E T^{xy}_E}$ with the operator $\del^2/\del k_z^2$ and take the limit $k_z \rightarrow 0$. This leaves out the integration over the Euclidean four-vector $q^\m$,
\bes
\int d^4q = \int^\infty_{-\infty} dq^0 \int^\infty_0 dq \,q^2 \int^\pi_0 d\th\, \text{sin}\,\th \int^{2\pi}_0 d\p,
\ees
where we have moved to spherical coordinates for the $\v[q]$ integral, with $q \equiv |\v[q]|$. The $\th, \p$ integrals are easy to perform and can be carried out first. The $q^0$ integral can then be performed using contour integration, noting that
\bes
\begin{split}
&\f{1}{\big((q^0 + i\m)^2 + q^2 + m^2\big)^n} = \f{(-1)^{n-1}}{2^{n-1} (n-1)!} \, \times \\ &\quad\underbrace{\f{1}{m}\f{\del}{\del m} \bigg( \f{1}{m}\f{\del}{\del m} \bigg(\cdots}_{n-1 \, \text{insertions of}\, \f{1}{m} \f{\del}{\del m}} \bigg( \f{1}{\big((q^0 + i\m)^2 + q^2 + m^2\big)}\bigg)\bigg)\bigg).
\end{split}
\ees
Thus, whenever the denominator of the integrand for the $q^0$ integral involves powers of $\big((q^0 + i\m)^2 + q^2 + m^2\big)$ greater than 1, we can rewrite the integrand with appropriate number of derivatives $\f{1}{m} \f{\del}{\del m}$ extracted out. The left-over $q^0$ integrand thus only has simple poles at $q^0 = -i (\m \pm \sqrt{q^2 + m^2})$, and the integral can be performed easily. Both the poles for the $q^0$ integral lie on the imaginary axis. For $\m > 0$, to have a non-zero contribution we must have $\m-\sqrt{q^2 + m^2} < 0$. On the other hand, for $\m < 0$ we must have $\m+\sqrt{q^2 + m^2} > 0$ to get a non-vanishing result. Both the conditions essentially restrict the domain of integration for $q$ to be $\sqrt{\m^2-m^2} \leq q \le \infty$, along with $|\m| > m$ for a non-zero result. 

We next perform the $q$ integral, which requires a UV cutoff $\L$ to regulate the divergences coming from short distance physics. The cutoff dependence can be removed systematically using the standard renormalization procedure of zero-temperature zero-density quantum field theory. Here we are only interested in $\m$ dependent terms, which appear independently from cutoff dependent terms. Keeping therefore only the $\m$ dependent terms and acting upon the result of the $q$ integration with appropriate factors of $\f{1}{m} \f{\del}{\del m}$ extracted out earlier, we get
\be
f_1 = - \f{1}{48\pi^2} \left[ |\m|\sqrt{\m^2 -m^2} - m^2 \log\left(\f{|\m|+\sqrt{\m^2 - m^2}}{m}\right)\right] \label{f1}
\ee
for $|\m| > m$, and $f_1  = 0$ for $|\m| < m$. Similar computations using the Kubo formulae eqs.\ \eqref{Kubof2}, \eqref{Kubof3} and eqs. \eqref{Kubof4} - \eqref{Kubof7} give the remaining second order susceptibilities for $|\m| > m$ to be
\begin{align} 
&f_2 = \f{|\m|}{24\pi^2} \f{2m^2-3\m^2}{\sqrt{\m^2 - m^2}}, \label{f2}\\
&f_3 = - \f{|\m|}{96\pi^2} \sqrt{\m^2 -m^2}, \label{f3}\\
&f_4 = \f{1}{12 \pi^2} \log\left(\f{|\m| + \sqrt{\m^2 - m^2}}{m}\right), \label{f4} \\
&f_5  = \f{\text{Sign}(\m)} {24 \pi^2} \sqrt{\m^2 -m^2}, \label{f5} \\
&f_6 = - \f{1}{24 \pi^2} \left[\f{|\m|}{\sqrt{\m^2 - m^2}} + 2 \log\left(\f{|\m|+\sqrt{\m^2 - m^2}}{m}\right)\right],\label{f6} \\
&f_7 = \f{\text{Sign}(\m)}{12\pi^2} \, \f{3\m^2 - 2m^2}{\sqrt{\m^2 - m^2}}. \label{f7}
\end{align}
Like $f_1$, the susceptibilities $f_2, \cdots f_7$ vanish for $|\m| < m$. Eqs.\ \eqref{f1} - \eqref{f7} are some of the main results of this paper. Note that $f_1, \cdots f_4$ and $f_6$ are even under charge conjugation $\m \rightarrow -\m$, whereas $f_5, f_7$ are odd.

An interesting limit to consider for the above results is the conformal limit, $m \rightarrow 0$.\footnote{See \cite{Buzzegoli:2018wpy} for a discussion with an axial chemical potential in the conformal limit.} In the conformal limit, the generating functional eq.\ \eqref{gfsecord} has to be invariant under a Weyl rescaling $g_{\m\n} \rightarrow \tilde{g}_{\m\n} = e^{-2 \xi} g_{\m\n}$ of the background metric to ensure the tracelessness of the energy-momentum tensor. A quantity $\chi$ that transforms under the above Weyl rescaling as $\chi \rightarrow \tilde{\chi} = e^{w \xi} \chi$ is said to have a Weyl weight $w$. Clearly the measure in eq.\ \eqref{gfsecord} has $w = - 4$, as $\sqrt{-g} \rightarrow e^{-4\xi} \sqrt{-g}$ under the Weyl rescaling. Thus, for the Weyl invariance of the generating functional eq.\ \eqref{gfsecord}, each of the terms $f_n s_n^{(2)}$ must have a Weyl weight of 4. The chemical potential $\m$ and the temperature $T$, defined through eq.\ \eqref{fluvar}, each have $w = 1$. The parity even quantities $\O^2$ and $B\cdot\O$ have well defined Weyl weights of $w = 2,3$ respectively. In the conformal limit, the Weyl weights of $f_3$ and $f_5$ from eqs. \eqref{f3} and \eqref{f5} are $w = 2, 1$ respectively, ensuring that the contribution of these two terms to the generating functional is Weyl invariant. 

The parity even terms $R, a^2$ and $E\cdot a$ do not have well defined Weyl weights. However, the combination
\bes
\int d^4x \sqrt{-g} \left( f\big(R+6a^2\big) - 6\, \f{\del f}{\del \m} E\cdot a\right)
\ees
is Weyl invariant at $T = 0$ for $f =  C\m^2$, where $C$ is a constant. Thus, for the Weyl invariance of the generating functional we expect that in the conformal limit $f_1 = C \m^2, f_2  = 6 f_1$ and $f_7 = - \, 6\, \del f_1/\del\m$. From eqs.\ \eqref{f1}, \eqref{f2} and \eqref{f7} we see that in the conformal limit $m\rightarrow 0$ we have $f_1 = - \m^2/48\pi^2, f_2 = - \m^2/8\pi^2$ and $f_7 = \m/4\pi^2$, thereby satisfying the criteria for Weyl invariance, with $C = - 1/48\pi^2$.

Interestingly, when a conformal field theory is coupled to a background metric and/or a background gauge field, the quantum corrections destroy the tracelessness of the theory. The ensuing trace anomaly is given by \cite{Duff:1993wm, Eling:2013bj}
\be
\begin{split}
\label{tanom}
g_{\m\n} T^{\m\n}& = -\f{a}{16\pi^2}\left( R_{\m\n\a\b}^{2} - 4R_{\m\n}^2 + R^2\right) \\
&+ \f{c}{16\pi^2} \left( R_{\mu\nu\a\b}^2 - 2R_{\m\n}^2 +\coeff13 R^2\right) - \f{b_0}{4} F_{\m\n}^2.
\end{split}
\ee
Here $a, c$ are numbers that depend upon the field content of the theory; for the theory of free massless Dirac fermions we have $a = 11/360, c = 1/20$. However, as the terms multiplying $a, c$ in the trace anomaly eq.\ \eqref{tanom} are fourth order in derivatives, they are not relevant for our present discussion as we are working only upto the second order. The last term in eq.\ \eqref{tanom} is however second order in derivatives, and must be encoded in the generating functional. Here $b_0$ is the coefficient of the one loop $\b$-function of the coupling $e$ used to minimally couple the theory to external electromagnetic fields,
\bes
M \f{d}{d M} \left(\f{1}{e^2}\right) = - \, b_0 + \mc{O}(e^2),
\ees
where $M$ is the renormalization scale, and the action for the external gauge fields is $- \f{1}{4 e^2} \int F_{\m\n}^2$.\footnote{In the limit $e \rightarrow 0$ the gauge field becomes non-dynamical. See section 2.1  of \cite{Fuini:2015hba} for an interesting discussion.} For a Dirac fermion we have $b_0 = 1/6\pi^2$. 

Consider now the terms proportional to $E^2, B^2$ in the generating functional, $\int d^4x \sqrt{-g}\left( f_4 B^2 + f_6 E^2\right)$, which on variation with respect to the metric give the following contribution to the trace of the energy-momentum tensor,
\be
g_{\m\n} T^{\m\n}_{f_4, f_6} = - \left(f'_4 B^2 + f'_6 E^2\right),
\ee
where $f'_n = \m \,\del_\m f_n + T\, \del_T f_n$. From eq.\ \eqref{tanom} this should be equal to $- \f{1}{24 \pi^2} F_{\m\n}^2 = \f{1}{12\pi^2} (E^2 - B^2)$, which implies that we should have
\be
f'_4 = 1/12 \pi^2\, , \quad f'_6 = - 1/12\pi^2.
\ee
This is indeed borne out by our results in eqs.\ \eqref{f4} and \eqref{f6}. Thus the susceptibilities are well behaved in the conformal limit, including the contribution to the trace anomaly. 

\section{Discussion}
\label{discussion}
We have computed the seven parity even equilibrium thermodynamic susceptibilities appearing at the second order in the hydrodynamic derivative expansion for a free massive Dirac field at zero-temperature and non-zero chemical potential. The resulting susceptibilities, eqs.\ \eqref{f1} - \eqref{f7}, can be inserted back into eqs.\ \eqref{constrel} and \eqref{constrelJ} to derive the constitutive relations for the Dirac field on a curved background. For the energy-momentum tensor the constitutive relations, with $\l \equiv \m/m > 1$, are\footnote{We write the constitutive relations for $\m > 0$. The results for $\m < 0$ can be obtained similarly using the appropriate signs in eqs.\ \eqref{f1} - \eqref{f7}.}
\begin{widetext}
\begin{subequations}
\label{constT}
\begin{align}
&\mc{E} = \e - \f{m^2}{48\pi^2}\left( \l\sqrt{\l^2 - 1} + \log\left(\l+\sqrt{\l^2 - 1}\right)\right) R + \f{m^2}{24\pi^2} \f{\l^3(3\l^2 - 4)}{(\l^2 - 1)^{3/2}}\,a^2 + \f{m^2}{32\pi^2} \f{\l (3\l^2 - 2)}{\sqrt{\l^2 - 1}}\,\O^2 \nonumber\\
&\hspace{77mm}+ \f{m^2}{24\pi^2} \left(\f{\l(1-3\l^2)}{\sqrt{\l^2 - 1}} - \log\left(\l+ \sqrt{\l^2 - 1}\right)\right) u^\a R_{\a\b} u^\b,\\
&\mc{P} = p - \f{m^2}{144\pi^2}\left(\l\sqrt{\l^2 - 1} - \log\left(\l+\sqrt{\l^2 - 1}\right)\right) R + \f{m^2}{24\pi^2} \f{\l^3}{\sqrt{\l^2 - 1}}\,a^2 + \f{m^2}{32\pi^2} \l\sqrt{\l^2 - 1}\, \O^2 \nonumber\\
&\hspace{77mm}- \f{m^2}{72\pi^2} \left( 3\l \sqrt{\l^2 - 1} + \log\left(\l+ \sqrt{\l^2 - 1}\right)\right) u^\a R_{\a\b} u^\b ,\\
&\mc{Q}_\m = - \f{m^2}{48\pi^2} \f{\l(4\l^2 - 3)}{\sqrt{\l^2-1}} \, \e_{\m\n\a\b} a^\n u^\a \O^\b - \f{m^2}{24\pi^2} \left( 2\l\sqrt{\l^2-1} - \log\left(\l+ \sqrt{\l^2 - 1}\right)\right)\,\D_\m^{\,\n} R_{\n\s} u^\s,\\
&\t_{\m\n} = - \f{m^2}{12\pi^2}\, \l\sqrt{\l^2-1}\, u^\a R_{\a\langle\m\n\rangle\b} u^\b + \f{m^2}{24\pi^2} \left( \l\sqrt{\l^2-1} - \log\left(\l+ \sqrt{\l^2 - 1}\right)\right) R_{\langle\m\n\rangle}.
\end{align}
\end{subequations}
\end{widetext}
Here $\e, p$ are zeroth-order energy density and pressure, eqs.\ \eqref{energy} and \eqref{pressure}. The constitutive relations for the current are
\begin{subequations}
\label{constJ2}
\begin{align}
&\mc{N} = n - \f{m}{24\pi^2} \sqrt{\l^2-1} \,R + \f{m}{24\pi^2}\f{6\l^4-9\l^2+2}{(\l^2-1)^{3/2}}\,a^2 \nonumber\\
&\quad+\f{m}{32\pi^2} \f{2\l^2-1}{\sqrt{\l^2-1}} \,\O^2 + \f{m}{12\pi^2} \f{(2-3\l^2)}{\sqrt{\l^2-1}}\, u^\a R_{\a\b} u^\b,\\
&\mc{J}^\m = -\f{m}{24\pi^2} \f{2\l^2-1}{\sqrt{\l^2-1}}\, \e^{\m\n\a\b} u_\n a_\a \O_\b \nonumber\\
&\hspace{30mm}+ \f{m}{12\pi^2} \sqrt{\l^2-1}\, \D^{\m\n} R_{\n\s} u^\s,
\end{align}
\end{subequations}
with $n$ denoting the zeroth-order charge density, eq.\ \eqref{numbdensity}. The coefficients of the two-derivative terms appearing in the constitutive relations above are the transport coefficients of the theory. 

The susceptibilities $f_2, f_6$ and $f_7$ diverge when the magnitude of the chemical potential approaches $m$ from above, i.e.\ in the limit $|\m| \rightarrow m^+$. This behaviour is carried on to some of the transport coefficients appearing in the constitutive relations eqs.\ \eqref{constT} and \eqref{constJ2}. As discussed in appendix \ref{FTE}, the apparent discontinuity across $|\m|/m = 1$ is a consequence of working strictly at zero temperature. At non-zero temperatures the susceptibilities are continuous functions of $\m/m$.

\iffalse
Note that in the limit $\m \rightarrow m^+$ the particle number density tends to zero,  eq.\ \eqref{numbdensity}. Thus we are dealing with an extremely rarefied system, in which the typical mean free paths of the particles are very large. This may also potentially affect the susceptibilities.
\fi

Second order thermodynamic susceptibilities for a free Dirac field were also computed in \cite{Megias:2014mba}. However, the results are for a massless field at finite temperature, rather than the massive case we have considered. Ref. \cite{Buzzegoli:2017cqy} computes the second order constitutive relations using equilibrium three-point functions, which are considerably harder to compute as opposed to the two-point functions we have used. As already mentioned, in equilibrium, the susceptibilities appearing in the generating functional are more fundamental than the transport coefficients in the constitutive relations, which are linear functions of the susceptibilities and their derivatives. Thus it is better to evaluate the susceptibilities first and then compute the transport coefficients from them, a route we have chosen. Besides this, the constitutive relations eqs.\ \eqref{constT} and \eqref{constJ2} exhibit explicit dependence on the background curvature, something not shown in \cite{Buzzegoli:2017cqy}. 

It would be interesting to make use of the thermodynamic susceptibilities computed here in a physical situation where second order derivative corrections may become relevant. Ref.\ \cite{Kovtun:2019wjz} for instance discusses the situation where a non-zero value for the susceptibilities can lead to a shift in the effective Newton's constant in the presence of matter.

\appendix
\section{Effects of non-zero temperature}
\label{FTE}
\begin{figure}[ht]
\includegraphics[scale=0.85]{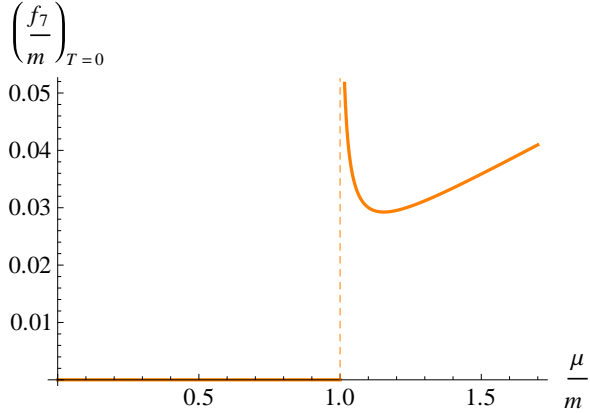}
\caption{Plot of $f_7$ as a function of $\m/m$ for $\m>0$ at zero temperature, eq.\ \eqref{f7}. There is a divergent discontinuity at $\m/m = 1$. Similar discontinuities also exist for $f_2$ and $f_6$, eqs.\ \eqref{f2} and \eqref{f6}.}
\label{f7zeroT}
\end{figure}

In this paper, we have computed parity-even second order equilibrium thermodynamic susceptibilities for a massive Dirac field at zero temperature, eqs.\ \eqref{f1}-\eqref{f7}. It seems from the results that the susceptibilities $f_2, f_6$ and $f_7$ have a divergent discontinuity across $|\m|/m = 1$, figure \ref{f7zeroT}. However, as we argue below, this discontinuity is an artifact of strictly imposing the $T \rightarrow 0$ limit. At any non-zero value of the temperature the susceptibilities are continuous and well-behaved.

To illustrate the point, let us consider the computation of $f_7$ at a non-zero temperature. The Kubo formula for $f_7$ is given by eq.\ \eqref{Kubof7}. Using the expressions for the energy-momentum tensor and the current presented in section \ref{Diracsetup}, the zero temperature expression for $f_7$ turns out to be
\be
\begin{split}
&f_7^{T=0} = - \half \lim_{\v[k]\to0} \frac{\partial^2}{\partial k_z^2} \int \f{d^4q}{(2\pi)^4} \\
&\quad\f{4i\tilde{q}_0\left(-\tilde{q}_0^2 + \v[q]^2 + m^2 -k_z q_z\right) + 8i\tilde{q}_0 q_x^2}{\left(\tilde{q}_0^2 + \v[q]^2 + m^2\right)\left(\tilde{q}_0^2 + \v[q]^2 + m^2 + k_z^2 - 2 k_z q_z\right)},
\label{f70T}
\end{split}
\ee
where $\tilde{q}_0$ is the shorthand for $q_0 + i\m$. When $T \neq 0$, the continuous integral over $q_0$ in the expression above has to be replaced by a discrete Matsubara sum,
\bes
q_0 \rightarrow \o_n = \f{(2n+1)\pi}{\b},\quad  \int^\infty_{-\infty} \f{dq_0}{2\pi} \rightarrow \f{1}{\b} \sum^\infty_{n=-\infty},
\ees
where $\b \equiv 1/T$ is the inverse temperature. Inserting the above in eq.\ \eqref{f70T}, performing the derivatives and integrating over the angular coordinates in $\v[q]$-space gives 
\be
f_7^{T\neq 0} = \f{1}{15\pi^2 \b} \int^\infty_0 dq \, q^2  \sum^{\infty}_{n=-\infty} \f{g(z)}{\left(z^2 - q^2 -m^2\right)^4}
\label{f7T1}
\ee
with
\bes
g(z) = 2z\left( 7q^4 +10q^2 \left( 3m^2 -4z^2\right)+15\left(m^4- z^4\right)\right),
\ees
where $q \equiv |\v[q]|$ as before, and $z =i \o_n - \m$.

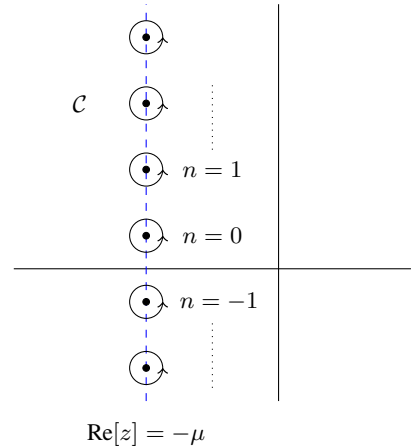
\begin{figure}[b]
\centering
\begin{tikzpicture}[scale=0.88]
\draw (-2,0) --(4,0);
\draw (2,-2) --(2,4);
\draw [blue, thin, dashed] (0,-2) --(0,4);
\draw [fill] (0,-1.5) circle [radius=0.05];
\draw [fill] (0,-0.5) circle [radius=0.05];
\draw [fill] (0,0.5) circle [radius=0.05];
\draw [fill] (0,1.5) circle [radius=0.05];
\draw [fill] (0,2.5) circle [radius=0.05];
\draw [fill] (0,3.5) circle [radius=0.05];
\draw [decoration={markings,mark=at position 0 with {\arrow[line width=0.2mm]{>}}},postaction={decorate}] (0,-1.5) circle [radius=0.25];
\draw [decoration={markings,mark=at position 0 with {\arrow[line width=0.2mm]{>}}},postaction={decorate}] (0,-0.5) circle [radius=0.25];
\draw [decoration={markings,mark=at position 0 with {\arrow[line width=0.2mm]{>}}},postaction={decorate}] (0,0.5) circle [radius=0.25];
\draw [decoration={markings,mark=at position 0 with {\arrow[line width=0.2mm]{>}}},postaction={decorate}] (0,1.5) circle [radius=0.25];
\draw [decoration={markings,mark=at position 0 with {\arrow[line width=0.2mm]{>}}},postaction={decorate}] (0,2.5) circle [radius=0.25];
\draw [decoration={markings,mark=at position 0 with {\arrow[line width=0.2mm]{>}}},postaction={decorate}] (0,3.5) circle [radius=0.25];
\node at (0,-2.5) {\small{$\text{Re}[z] =  -\m$}};
\node at (-1.0,2.5) {$\mc{C}$};
\node at (1.1,-0.5) {$n=-1$};
\node at (1.0,0.5) {$n=0$};
\node at (1.0,1.5) {$n=1$};
\draw [black, dotted] (1,-1.8) --(1,-0.8);
\draw [black, dotted] (1,1.8) --(1,2.8);
\end{tikzpicture}
\caption{The contour $\mc{C}$ on the complex $z$ plane alluded to in eq.\eqref{f7T2}. The dots denote the points $z = i \o_n - \m$ for $n \in \mathbb{Z}$.}
\label{contour1}
\end{figure}

\begin{figure}[htp]
\centering
\begin{tikzpicture}[scale=0.88]
\draw (6,0) --(13.5,0);
\draw (11,-2) --(11,4);
\draw [blue, thin, dashed] (9.5,-2) --(9.5,4);
\draw [fill] (9.5,-1.5) circle [radius=0.04];
\draw [fill] (9.5,-0.5) circle [radius=0.04];
\draw [fill] (9.5,0.5) circle [radius=0.04];
\draw [fill] (9.5,1.5) circle [radius=0.04];
\draw [fill] (9.5,2.5) circle [radius=0.04];
\draw [fill] (9.5,3.5) circle [radius=0.04];
\draw [decoration={markings,mark=at position 1/2 with {\arrow[line width=0.5mm]{>}}},postaction={decorate}] [thick] (9.8,-2) --(9.8,4);
\draw [decoration={markings,mark=at position 1/2 with {\arrow[line width=0.5mm]{<}}},postaction={decorate}] [thick] (9.2,-2) --(9.2,4);
\node at (10.2,2) {$\mc{C}_1$};
\node at (8.9,2) {$\mc{C}_2$};
\draw [decoration={markings,mark=at position 1/2 with {\arrow[line width=0.4mm]{>}}},postaction={decorate}] [dashed, purple] (9.8,4) arc (90:-90:3);
\draw [decoration={markings,mark=at position 1/2 with {\arrow[line width=0.4mm]{<}}},postaction={decorate}] [dashed, purple] (9.2,4) arc (-90:90:-3);
\node at (13,2) {$\mc{C}_1'$};
\node at (6,2) {$\mc{C}_2'$};
\node at (11,-2.5) {\small{$\text{Re}[z] = -\m + \e$}};
\node at (8,-2.5) {\small{$\text{Re}[z] = -\m - \e$}};
\draw [thin, gray, ->] (8.4,-2.2)--(9.1,-1.5);
\draw [thin, gray, ->] (10.6,-2.2)--(9.9,-1.5);
\end{tikzpicture}
\caption{The contour $\mc{C}_1\cup\mc{C}_2$. The separation $2\e$ between $\mc{C}_1$ and $\mc{C}_2$ tends to zero. $\mc{C}_1'$ and $\mc{C}_2'$ are oriented semicircular arcs of $\infty$ radius which are used to close the contours $\mc{C}_1$ and $\mc{C}_2$, respectively.}
\label{contour2}
\end{figure}
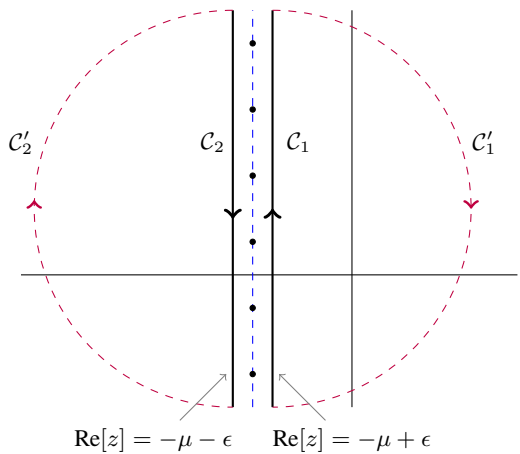

The next step in the evaluation of $f_7$ at non-zero temperatures involves evaluating the Matsubara sum in eq.\ \eqref{f7T1}. This can be done by converting the sum into a contour integral \cite{Kapusta:2006pm, Bellac:2011kqa}. Consider the function $\f{\b}{2}\, \text{tanh} \left( \f{\b(z+\m)}{2}\right)$. This function has poles at $z = i\o_n - \m$, each with a unit residue, and is otherwise well behaved on the entire complex plane, including being bounded for the limit $z \rightarrow \infty$. This allows us to write the infinite sum in eq.\ \eqref{f7T1} as a contour integral,
\be
\begin{split}
\mc{S} &= \sum^{\infty}_{n=-\infty} \f{g(z)}{\left(z^2 - q^2 -m^2\right)^4} \\
&= \int_{\mc{C}} \f{dz}{2\pi i}\, \f{g(z)}{\left(z^2 - q^2 -m^2\right)^4} \, \f{\b}{2}\, \text{tanh} \left( \f{\b(z+\m)}{2}\right),
\label{f7T2}
\end{split}
\ee
where the contour $\mc{C}$ is shown in figure \ref{contour1}. This contour can be continuously deformed into the contour $\mc{C}_1 \cup \mc{C}_2$, figure \ref{contour2}. Now, due to the fact that the integrand in eq.\ \eqref{f7T2} falls off faster than $1/z$ for $z \rightarrow \infty$, we can close $\mc{C}_1$ with a semicircle $\mc{C}_1'$ to its right and $\mc{C}_2$ with a semicircle $\mc{C}_2'$ to its left, without affecting the value of $\mc{S}$. From the residue theorem it is then clear that the contributions to $\mc{S}$ will only come from the poles at $z = \pm \sqrt{q^2 + m^2} \equiv z_\pm$, given by
\be
\mc{S} = - \sum_{z = z_\pm} \text{Res}\left[ \f{g(z)}{\left(z^2 - q^2 -m^2\right)^4} \, \f{\b}{2}\, \text{tanh} \left( \f{\b(z+\m)}{2}\right)\right],
\label{sum1}
\ee
with the minus sign coming because we have closed the contours in a  clockwise manner.
 
\begin{figure}[h]
\includegraphics[scale=0.90]{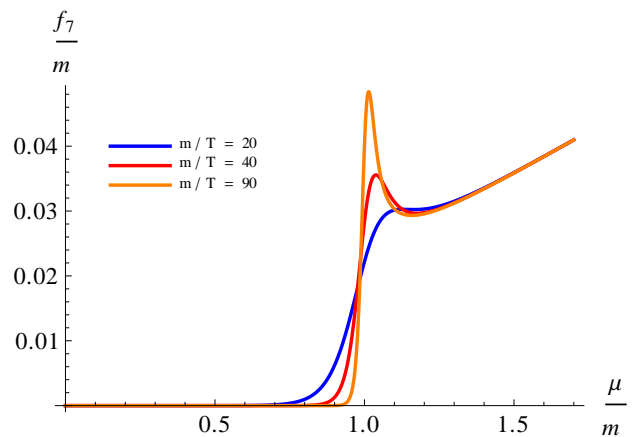}
\caption{Plots depicting the behaviour of $f_7$ as a function of $\m/m$ for $\m>0$ at different non-zero temperatures. As the temperature becomes smaller, the jump in $f_7$ as well as the peak height around $\m/m = 1$ become more prominent, with the limiting case $T = 0$ plotted in figure \ref{f7zeroT}. In an identical manner the non-zero temperature also removes the discontinuity at $\m = -m$.}
\label{f7T}
\end{figure}

The result in eq.\ \eqref{sum1} above can be further simplified by noting that
\bes
\text{tanh} \left( \f{\b(z+\m)}{2}\right) = 1 - 2\,\mc{F}_{-\m}(z),
\ees
where $\mc{F}_\m(z) = 1/\left(e^{\b(z-\m)}+1\right)$ is the Fermi-Dirac distribution function. Inserting this in eq.\ \eqref{sum1} gives
\be
\begin{split}
\mc{S} = &- \f{\b}{2} \sum_{z = z_\pm} \text{Res}\left[ \f{g(z)}{\left(z^2 - q^2 -m^2\right)^4}\right] \\
&+ \b \sum_{z = z_\pm} \text{Res}\left[ \f{g(z)\, \mc{F}_{-\m}(z)}{\left(z^2 - q^2 -m^2\right)^4}\right].
\end{split}
\ee
The first term in the expression above vanishes, leaving behind the second term, which when substituted into eq.\ \eqref{f7T1} gives
\be
f_7^{T\neq 0} = \f{1}{15\pi^2} \int^\infty_0 dq \, q^2  \sum_{z = z_\pm} \text{Res}\left[ \f{g(z)\, \mc{F}_{-\m}(z)}{\left(z^2 - q^2 -m^2\right)^4}\right].
\label{f7T3}
\ee
Although the residues in eq.\ \eqref{f7T3} above can be readily computed, the leftover integration on $q$ is difficult to perform. The resulting integrand however is a well-behaved  function of $q$, and for the sake of illustration, can be integrated numerically for different choices of the temperature $T$. The resulting behaviour of $f_7$ as a function of $\m/m$ is plotted in figure \ref{f7T} for $\m > 0$ and different choices of the temperature. As is evident from the figure, the smaller the temperature, the stronger the jump in $f_7$ around $\m/m = 1$, with the limiting case occurring at $T = 0$ as shown in figure \ref{f7zeroT}.    

The apparent discontinuities in $f_2$ and $f_6$ also smoothen out once one deviates from the strict zero temperature condition, just like the case for $f_7$. One has to be careful while performing the numerical integration step to obtain the plots for $f_2$ and $f_6$ at a non-zero temperature due to UV divergences, which have to be carefully subtracted off by adding the appropriate counter terms. The results for $f_2$ and $f_6$ then are also continuous and well behaved. 

\acknowledgments
The author would like to thank Pavel Kovtun for insightful discussions, and for comments on a draft of this paper which lead to several improvements. The work is supported by the NSERC of Canada.

\bibliography{DegFermi}

\begin{thebibliography}{30}%
\makeatletter
\providecommand \@ifxundefined [1]{%
 \@ifx{#1\undefined}
}%
\providecommand \@ifnum [1]{%
 \ifnum #1\expandafter \@firstoftwo
 \else \expandafter \@secondoftwo
 \fi
}%
\providecommand \@ifx [1]{%
 \ifx #1\expandafter \@firstoftwo
 \else \expandafter \@secondoftwo
 \fi
}%
\providecommand \natexlab [1]{#1}%
\providecommand \enquote  [1]{``#1''}%
\providecommand \bibnamefont  [1]{#1}%
\providecommand \bibfnamefont [1]{#1}%
\providecommand \citenamefont [1]{#1}%
\providecommand \href@noop [0]{\@secondoftwo}%
\providecommand \href [0]{\begingroup \@sanitize@url \@href}%
\providecommand \@href[1]{\@@startlink{#1}\@@href}%
\providecommand \@@href[1]{\endgroup#1\@@endlink}%
\providecommand \@sanitize@url [0]{\catcode `\\12\catcode `\$12\catcode
  `\&12\catcode `\#12\catcode `\^12\catcode `\_12\catcode `\%12\relax}%
\providecommand \@@startlink[1]{}%
\providecommand \@@endlink[0]{}%
\providecommand \url  [0]{\begingroup\@sanitize@url \@url }%
\providecommand \@url [1]{\endgroup\@href {#1}{\urlprefix }}%
\providecommand \urlprefix  [0]{URL }%
\providecommand \Eprint [0]{\href }%
\providecommand \doibase [0]{http://dx.doi.org/}%
\providecommand \selectlanguage [0]{\@gobble}%
\providecommand \bibinfo  [0]{\@secondoftwo}%
\providecommand \bibfield  [0]{\@secondoftwo}%
\providecommand \translation [1]{[#1]}%
\providecommand \BibitemOpen [0]{}%
\providecommand \bibitemStop [0]{}%
\providecommand \bibitemNoStop [0]{.\EOS\space}%
\providecommand \EOS [0]{\spacefactor3000\relax}%
\providecommand \BibitemShut  [1]{\csname bibitem#1\endcsname}%
\let\auto@bib@innerbib\@empty
%</preamble>
\bibitem [{\citenamefont {Kovtun}\ and\ \citenamefont
  {Shukla}(2018)}]{Kovtun:2018dvd}%
  \BibitemOpen
  \bibfield  {author} {\bibinfo {author} {\bibfnamefont {Pavel}\ \bibnamefont
  {Kovtun}}\ and\ \bibinfo {author} {\bibfnamefont {Ashish}\ \bibnamefont
  {Shukla}},\ }\bibfield  {title} {\enquote {\bibinfo {title} {{Kubo formulas
  for thermodynamic transport coefficients}},}\ }\href {\doibase
  10.1007/JHEP10(2018)007} {\bibfield  {journal} {\bibinfo  {journal} {JHEP}\
  }\textbf {\bibinfo {volume} {10}},\ \bibinfo {pages} {007} (\bibinfo {year}
  {2018})},\ \Eprint {http://arxiv.org/abs/1806.05774} {arXiv:1806.05774
  [hep-th]} \BibitemShut {NoStop}%
%%CITATION = ARXIV:1806.05774;%%
\bibitem [{\citenamefont {Landau}\ and\ \citenamefont
  {Lifshitz}(2013)}]{landau2013fluid}%
  \BibitemOpen
  \bibfield  {author} {\bibinfo {author} {\bibfnamefont {L.D.}\ \bibnamefont
  {Landau}}\ and\ \bibinfo {author} {\bibfnamefont {E.M.}\ \bibnamefont
  {Lifshitz}},\ }\href {https://books.google.ca/books?id=CeBbAwAAQBAJ} {\emph
  {\bibinfo {title} {Fluid Mechanics}}},\ \bibinfo {number} {v. 6}\ (\bibinfo
  {publisher} {Elsevier Science},\ \bibinfo {year} {2013})\BibitemShut
  {NoStop}%
\bibitem [{\citenamefont {Kovtun}(2012)}]{Kovtun:2012rj}%
  \BibitemOpen
  \bibfield  {author} {\bibinfo {author} {\bibfnamefont {Pavel}\ \bibnamefont
  {Kovtun}},\ }\bibfield  {title} {\enquote {\bibinfo {title} {{Lectures on
  hydrodynamic fluctuations in relativistic theories}},}\ }\bibfield
  {booktitle} {\emph {\bibinfo {booktitle} {{INT Summer School on Applications
  of String Theory Seattle, Washington, USA, July 18-29, 2011}}},\ }\href
  {\doibase 10.1088/1751-8113/45/47/473001} {\bibfield  {journal} {\bibinfo
  {journal} {J. Phys.}\ }\textbf {\bibinfo {volume} {A45}},\ \bibinfo {pages}
  {473001} (\bibinfo {year} {2012})},\ \Eprint {http://arxiv.org/abs/1205.5040}
  {arXiv:1205.5040 [hep-th]} \BibitemShut {NoStop}%
%%CITATION = ARXIV:1205.5040;%%
\bibitem [{\citenamefont {Romatschke}\ and\ \citenamefont
  {Romatschke}(2019)}]{Romatschke:2017ejr}%
  \BibitemOpen
  \bibfield  {author} {\bibinfo {author} {\bibfnamefont {Paul}\ \bibnamefont
  {Romatschke}}\ and\ \bibinfo {author} {\bibfnamefont {Ulrike}\ \bibnamefont
  {Romatschke}},\ }\bibfield  {title} {\enquote {\bibinfo {title}
  {{Relativistic Fluid Dynamics In and Out of Equilibrium}},}\ }\href@noop {}
  {\  (\bibinfo {year} {2019})},\ \Eprint {http://arxiv.org/abs/1712.05815}
  {arXiv:1712.05815 [nucl-th]} \BibitemShut {NoStop}%
%%CITATION = ARXIV:1712.05815;%%
\bibitem [{\citenamefont {Baier}\ \emph {et~al.}(2008)\citenamefont {Baier},
  \citenamefont {Romatschke}, \citenamefont {Son}, \citenamefont {Starinets},\
  and\ \citenamefont {Stephanov}}]{Baier:2007ix}%
  \BibitemOpen
  \bibfield  {author} {\bibinfo {author} {\bibfnamefont {Rudolf}\ \bibnamefont
  {Baier}}, \bibinfo {author} {\bibfnamefont {Paul}\ \bibnamefont
  {Romatschke}}, \bibinfo {author} {\bibfnamefont {Dam~Thanh}\ \bibnamefont
  {Son}}, \bibinfo {author} {\bibfnamefont {Andrei~O.}\ \bibnamefont
  {Starinets}}, \ and\ \bibinfo {author} {\bibfnamefont {Mikhail~A.}\
  \bibnamefont {Stephanov}},\ }\bibfield  {title} {\enquote {\bibinfo {title}
  {{Relativistic viscous hydrodynamics, conformal invariance, and
  holography}},}\ }\href {\doibase 10.1088/1126-6708/2008/04/100} {\bibfield
  {journal} {\bibinfo  {journal} {JHEP}\ }\textbf {\bibinfo {volume} {04}},\
  \bibinfo {pages} {100} (\bibinfo {year} {2008})},\ \Eprint
  {http://arxiv.org/abs/0712.2451} {arXiv:0712.2451 [hep-th]} \BibitemShut
  {NoStop}%
%%CITATION = ARXIV:0712.2451;%%
\bibitem [{\citenamefont {Bhattacharyya}\ \emph {et~al.}(2008)\citenamefont
  {Bhattacharyya}, \citenamefont {Hubeny}, \citenamefont {Minwalla},\ and\
  \citenamefont {Rangamani}}]{Bhattacharyya:2008jc}%
  \BibitemOpen
  \bibfield  {author} {\bibinfo {author} {\bibfnamefont {Sayantani}\
  \bibnamefont {Bhattacharyya}}, \bibinfo {author} {\bibfnamefont {Veronika~E}\
  \bibnamefont {Hubeny}}, \bibinfo {author} {\bibfnamefont {Shiraz}\
  \bibnamefont {Minwalla}}, \ and\ \bibinfo {author} {\bibfnamefont {Mukund}\
  \bibnamefont {Rangamani}},\ }\bibfield  {title} {\enquote {\bibinfo {title}
  {{Nonlinear Fluid Dynamics from Gravity}},}\ }\href {\doibase
  10.1088/1126-6708/2008/02/045} {\bibfield  {journal} {\bibinfo  {journal}
  {JHEP}\ }\textbf {\bibinfo {volume} {02}},\ \bibinfo {pages} {045} (\bibinfo
  {year} {2008})},\ \Eprint {http://arxiv.org/abs/0712.2456} {arXiv:0712.2456
  [hep-th]} \BibitemShut {NoStop}%
%%CITATION = ARXIV:0712.2456;%%
\bibitem [{\citenamefont {Jensen}\ \emph
  {et~al.}(2012{\natexlab{a}})\citenamefont {Jensen}, \citenamefont {Kaminski},
  \citenamefont {Kovtun}, \citenamefont {Meyer}, \citenamefont {Ritz},\ and\
  \citenamefont {Yarom}}]{Jensen:2011xb}%
  \BibitemOpen
  \bibfield  {author} {\bibinfo {author} {\bibfnamefont {Kristan}\ \bibnamefont
  {Jensen}}, \bibinfo {author} {\bibfnamefont {Matthias}\ \bibnamefont
  {Kaminski}}, \bibinfo {author} {\bibfnamefont {Pavel}\ \bibnamefont
  {Kovtun}}, \bibinfo {author} {\bibfnamefont {Rene}\ \bibnamefont {Meyer}},
  \bibinfo {author} {\bibfnamefont {Adam}\ \bibnamefont {Ritz}}, \ and\
  \bibinfo {author} {\bibfnamefont {Amos}\ \bibnamefont {Yarom}},\ }\bibfield
  {title} {\enquote {\bibinfo {title} {{Parity-Violating Hydrodynamics in 2+1
  Dimensions}},}\ }\href {\doibase 10.1007/JHEP05(2012)102} {\bibfield
  {journal} {\bibinfo  {journal} {JHEP}\ }\textbf {\bibinfo {volume} {05}},\
  \bibinfo {pages} {102} (\bibinfo {year} {2012}{\natexlab{a}})},\ \Eprint
  {http://arxiv.org/abs/1112.4498} {arXiv:1112.4498 [hep-th]} \BibitemShut
  {NoStop}%
%%CITATION = ARXIV:1112.4498;%%
\bibitem [{\citenamefont {Banerjee}\ \emph {et~al.}(2012)\citenamefont
  {Banerjee}, \citenamefont {Bhattacharya}, \citenamefont {Bhattacharyya},
  \citenamefont {Jain}, \citenamefont {Minwalla},\ and\ \citenamefont
  {Sharma}}]{Banerjee:2012iz}%
  \BibitemOpen
  \bibfield  {author} {\bibinfo {author} {\bibfnamefont {Nabamita}\
  \bibnamefont {Banerjee}}, \bibinfo {author} {\bibfnamefont {Jyotirmoy}\
  \bibnamefont {Bhattacharya}}, \bibinfo {author} {\bibfnamefont {Sayantani}\
  \bibnamefont {Bhattacharyya}}, \bibinfo {author} {\bibfnamefont {Sachin}\
  \bibnamefont {Jain}}, \bibinfo {author} {\bibfnamefont {Shiraz}\ \bibnamefont
  {Minwalla}}, \ and\ \bibinfo {author} {\bibfnamefont {Tarun}\ \bibnamefont
  {Sharma}},\ }\bibfield  {title} {\enquote {\bibinfo {title} {{Constraints on
  Fluid Dynamics from Equilibrium Partition Functions}},}\ }\href {\doibase
  10.1007/JHEP09(2012)046} {\bibfield  {journal} {\bibinfo  {journal} {JHEP}\
  }\textbf {\bibinfo {volume} {09}},\ \bibinfo {pages} {046} (\bibinfo {year}
  {2012})},\ \Eprint {http://arxiv.org/abs/1203.3544} {arXiv:1203.3544
  [hep-th]} \BibitemShut {NoStop}%
%%CITATION = ARXIV:1203.3544;%%
\bibitem [{\citenamefont {Jensen}\ \emph
  {et~al.}(2012{\natexlab{b}})\citenamefont {Jensen}, \citenamefont {Kaminski},
  \citenamefont {Kovtun}, \citenamefont {Meyer}, \citenamefont {Ritz},\ and\
  \citenamefont {Yarom}}]{Jensen:2012jh}%
  \BibitemOpen
  \bibfield  {author} {\bibinfo {author} {\bibfnamefont {Kristan}\ \bibnamefont
  {Jensen}}, \bibinfo {author} {\bibfnamefont {Matthias}\ \bibnamefont
  {Kaminski}}, \bibinfo {author} {\bibfnamefont {Pavel}\ \bibnamefont
  {Kovtun}}, \bibinfo {author} {\bibfnamefont {Rene}\ \bibnamefont {Meyer}},
  \bibinfo {author} {\bibfnamefont {Adam}\ \bibnamefont {Ritz}}, \ and\
  \bibinfo {author} {\bibfnamefont {Amos}\ \bibnamefont {Yarom}},\ }\bibfield
  {title} {\enquote {\bibinfo {title} {{Towards hydrodynamics without an
  entropy current}},}\ }\href {\doibase 10.1103/PhysRevLett.109.101601}
  {\bibfield  {journal} {\bibinfo  {journal} {Phys. Rev. Lett.}\ }\textbf
  {\bibinfo {volume} {109}},\ \bibinfo {pages} {101601} (\bibinfo {year}
  {2012}{\natexlab{b}})},\ \Eprint {http://arxiv.org/abs/1203.3556}
  {arXiv:1203.3556 [hep-th]} \BibitemShut {NoStop}%
%%CITATION = ARXIV:1203.3556;%%
\bibitem [{\citenamefont {Shapiro}\ and\ \citenamefont
  {Teukolsky}(1983)}]{Shapiro:1983du}%
  \BibitemOpen
  \bibfield  {author} {\bibinfo {author} {\bibfnamefont {S.~L.}\ \bibnamefont
  {Shapiro}}\ and\ \bibinfo {author} {\bibfnamefont {S.~A.}\ \bibnamefont
  {Teukolsky}},\ }\href@noop {} {\emph {\bibinfo {title} {{Black holes, white
  dwarfs, and neutron stars: The physics of compact objects}}}}\ (\bibinfo
  {year} {1983})\BibitemShut {NoStop}%
%%CITATION = INSPIRE-198204;%%
\bibitem [{\citenamefont {Glendenning}(1997)}]{Glendenning:1997wn}%
  \BibitemOpen
  \bibfield  {author} {\bibinfo {author} {\bibfnamefont {N.~K.}\ \bibnamefont
  {Glendenning}},\ }\href@noop {} {\emph {\bibinfo {title} {{Compact stars:
  Nuclear physics, particle physics, and general relativity}}}}\ (\bibinfo
  {year} {1997})\BibitemShut {NoStop}%
%%CITATION = INSPIRE-456851;%%
\bibitem [{\citenamefont {Vuorinen}(2019)}]{Vuorinen:2018qzx}%
  \BibitemOpen
  \bibfield  {author} {\bibinfo {author} {\bibfnamefont {Aleksi}\ \bibnamefont
  {Vuorinen}},\ }\bibfield  {title} {\enquote {\bibinfo {title} {{Neutron stars
  and stellar mergers as a laboratory for dense QCD matter}},}\ }\bibfield
  {booktitle} {\emph {\bibinfo {booktitle} {{Proceedings, 27th International
  Conference on Ultrarelativistic Nucleus-Nucleus Collisions (Quark Matter
  2018): Venice, Italy, May 14-19, 2018}}},\ }\href {\doibase
  10.1016/j.nuclphysa.2018.10.011} {\bibfield  {journal} {\bibinfo  {journal}
  {Nucl. Phys.}\ }\textbf {\bibinfo {volume} {A982}},\ \bibinfo {pages}
  {36--42} (\bibinfo {year} {2019})},\ \Eprint
  {http://arxiv.org/abs/1807.04480} {arXiv:1807.04480 [nucl-th]} \BibitemShut
  {NoStop}%
%%CITATION = ARXIV:1807.04480;%%
\bibitem [{\citenamefont {Annala}\ \emph {et~al.}(2019)\citenamefont {Annala},
  \citenamefont {Gorda}, \citenamefont {Kurkela}, \citenamefont {Nättilä},\
  and\ \citenamefont {Vuorinen}}]{Annala:2019eax}%
  \BibitemOpen
  \bibfield  {author} {\bibinfo {author} {\bibfnamefont {Eemeli}\ \bibnamefont
  {Annala}}, \bibinfo {author} {\bibfnamefont {Tyler}\ \bibnamefont {Gorda}},
  \bibinfo {author} {\bibfnamefont {Aleksi}\ \bibnamefont {Kurkela}}, \bibinfo
  {author} {\bibfnamefont {Joonas}\ \bibnamefont {Nättilä}}, \ and\ \bibinfo
  {author} {\bibfnamefont {Aleksi}\ \bibnamefont {Vuorinen}},\ }\bibfield
  {title} {\enquote {\bibinfo {title} {{Constraining the properties of
  neutron-star matter with observations}},}\ }in\ \href@noop {} {\emph
  {\bibinfo {booktitle} {{12th INTEGRAL conference and 1st AHEAD Gamma-ray
  workshop (INTEGRAL 2019): INTEGRAL looks AHEAD to Multi-Messenger
  Astrophysics Geneva, Switzerland, February 11-15, 2019}}}}\ (\bibinfo {year}
  {2019})\ \Eprint {http://arxiv.org/abs/1904.01354} {arXiv:1904.01354
  [astro-ph.HE]} \BibitemShut {NoStop}%
%%CITATION = ARXIV:1904.01354;%%
\bibitem [{\citenamefont {Alford}\ \emph {et~al.}(2008)\citenamefont {Alford},
  \citenamefont {Schmitt}, \citenamefont {Rajagopal},\ and\ \citenamefont
  {Schäfer}}]{Alford:2007xm}%
  \BibitemOpen
  \bibfield  {author} {\bibinfo {author} {\bibfnamefont {Mark~G.}\ \bibnamefont
  {Alford}}, \bibinfo {author} {\bibfnamefont {Andreas}\ \bibnamefont
  {Schmitt}}, \bibinfo {author} {\bibfnamefont {Krishna}\ \bibnamefont
  {Rajagopal}}, \ and\ \bibinfo {author} {\bibfnamefont {Thomas}\ \bibnamefont
  {Schäfer}},\ }\bibfield  {title} {\enquote {\bibinfo {title} {{Color
  superconductivity in dense quark matter}},}\ }\href {\doibase
  10.1103/RevModPhys.80.1455} {\bibfield  {journal} {\bibinfo  {journal} {Rev.
  Mod. Phys.}\ }\textbf {\bibinfo {volume} {80}},\ \bibinfo {pages}
  {1455--1515} (\bibinfo {year} {2008})},\ \Eprint
  {http://arxiv.org/abs/0709.4635} {arXiv:0709.4635 [hep-ph]} \BibitemShut
  {NoStop}%
%%CITATION = ARXIV:0709.4635;%%
\bibitem [{\citenamefont {Kurkela}\ \emph {et~al.}(2010)\citenamefont
  {Kurkela}, \citenamefont {Romatschke},\ and\ \citenamefont
  {Vuorinen}}]{Kurkela:2009gj}%
  \BibitemOpen
  \bibfield  {author} {\bibinfo {author} {\bibfnamefont {Aleksi}\ \bibnamefont
  {Kurkela}}, \bibinfo {author} {\bibfnamefont {Paul}\ \bibnamefont
  {Romatschke}}, \ and\ \bibinfo {author} {\bibfnamefont {Aleksi}\ \bibnamefont
  {Vuorinen}},\ }\bibfield  {title} {\enquote {\bibinfo {title} {{Cold Quark
  Matter}},}\ }\href {\doibase 10.1103/PhysRevD.81.105021} {\bibfield
  {journal} {\bibinfo  {journal} {Phys. Rev.}\ }\textbf {\bibinfo {volume}
  {D81}},\ \bibinfo {pages} {105021} (\bibinfo {year} {2010})},\ \Eprint
  {http://arxiv.org/abs/0912.1856} {arXiv:0912.1856 [hep-ph]} \BibitemShut
  {NoStop}%
%%CITATION = ARXIV:0912.1856;%%
\bibitem [{\citenamefont {Ghisoiu}\ \emph {et~al.}(2017)\citenamefont
  {Ghisoiu}, \citenamefont {Gorda}, \citenamefont {Kurkela}, \citenamefont
  {Romatschke}, \citenamefont {Säppi},\ and\ \citenamefont
  {Vuorinen}}]{Ghisoiu:2016swa}%
  \BibitemOpen
  \bibfield  {author} {\bibinfo {author} {\bibfnamefont {Ioan}\ \bibnamefont
  {Ghisoiu}}, \bibinfo {author} {\bibfnamefont {Tyler}\ \bibnamefont {Gorda}},
  \bibinfo {author} {\bibfnamefont {Aleksi}\ \bibnamefont {Kurkela}}, \bibinfo
  {author} {\bibfnamefont {Paul}\ \bibnamefont {Romatschke}}, \bibinfo {author}
  {\bibfnamefont {Matias}\ \bibnamefont {Säppi}}, \ and\ \bibinfo {author}
  {\bibfnamefont {Aleksi}\ \bibnamefont {Vuorinen}},\ }\bibfield  {title}
  {\enquote {\bibinfo {title} {{On high-order perturbative calculations at
  finite density}},}\ }\href {\doibase 10.1016/j.nuclphysb.2016.11.023}
  {\bibfield  {journal} {\bibinfo  {journal} {Nucl. Phys.}\ }\textbf {\bibinfo
  {volume} {B915}},\ \bibinfo {pages} {102--118} (\bibinfo {year} {2017})},\
  \Eprint {http://arxiv.org/abs/1609.04339} {arXiv:1609.04339 [hep-ph]}
  \BibitemShut {NoStop}%
%%CITATION = ARXIV:1609.04339;%%
\bibitem [{\citenamefont {Kovtun}(2016)}]{Kovtun:2016lfw}%
  \BibitemOpen
  \bibfield  {author} {\bibinfo {author} {\bibfnamefont {Pavel}\ \bibnamefont
  {Kovtun}},\ }\bibfield  {title} {\enquote {\bibinfo {title} {{Thermodynamics
  of polarized relativistic matter}},}\ }\href {\doibase
  10.1007/JHEP07(2016)028} {\bibfield  {journal} {\bibinfo  {journal} {JHEP}\
  }\textbf {\bibinfo {volume} {07}},\ \bibinfo {pages} {028} (\bibinfo {year}
  {2016})},\ \Eprint {http://arxiv.org/abs/1606.01226} {arXiv:1606.01226
  [hep-th]} \BibitemShut {NoStop}%
%%CITATION = ARXIV:1606.01226;%%
\bibitem [{\citenamefont {Hernandez}\ and\ \citenamefont
  {Kovtun}(2017)}]{Hernandez:2017mch}%
  \BibitemOpen
  \bibfield  {author} {\bibinfo {author} {\bibfnamefont {Juan}\ \bibnamefont
  {Hernandez}}\ and\ \bibinfo {author} {\bibfnamefont {Pavel}\ \bibnamefont
  {Kovtun}},\ }\bibfield  {title} {\enquote {\bibinfo {title} {{Relativistic
  magnetohydrodynamics}},}\ }\href {\doibase 10.1007/JHEP05(2017)001}
  {\bibfield  {journal} {\bibinfo  {journal} {JHEP}\ }\textbf {\bibinfo
  {volume} {05}},\ \bibinfo {pages} {001} (\bibinfo {year} {2017})},\ \Eprint
  {http://arxiv.org/abs/1703.08757} {arXiv:1703.08757 [hep-th]} \BibitemShut
  {NoStop}%
%%CITATION = ARXIV:1703.08757;%%
\bibitem [{\citenamefont {Moore}\ and\ \citenamefont
  {Sohrabi}(2011)}]{Moore:2010bu}%
  \BibitemOpen
  \bibfield  {author} {\bibinfo {author} {\bibfnamefont {Guy~D.}\ \bibnamefont
  {Moore}}\ and\ \bibinfo {author} {\bibfnamefont {Kiyoumars~A.}\ \bibnamefont
  {Sohrabi}},\ }\bibfield  {title} {\enquote {\bibinfo {title} {{Kubo Formulae
  for Second-Order Hydrodynamic Coefficients}},}\ }\href {\doibase
  10.1103/PhysRevLett.106.122302} {\bibfield  {journal} {\bibinfo  {journal}
  {Phys. Rev. Lett.}\ }\textbf {\bibinfo {volume} {106}},\ \bibinfo {pages}
  {122302} (\bibinfo {year} {2011})},\ \Eprint {http://arxiv.org/abs/1007.5333}
  {arXiv:1007.5333 [hep-ph]} \BibitemShut {NoStop}%
%%CITATION = ARXIV:1007.5333;%%
\bibitem [{\citenamefont {Moore}\ and\ \citenamefont
  {Sohrabi}(2012)}]{Moore:2012tc}%
  \BibitemOpen
  \bibfield  {author} {\bibinfo {author} {\bibfnamefont {Guy~D.}\ \bibnamefont
  {Moore}}\ and\ \bibinfo {author} {\bibfnamefont {Kiyoumars~A.}\ \bibnamefont
  {Sohrabi}},\ }\bibfield  {title} {\enquote {\bibinfo {title}
  {{Thermodynamical second-order hydrodynamic coefficients}},}\ }\href
  {\doibase 10.1007/JHEP11(2012)148} {\bibfield  {journal} {\bibinfo  {journal}
  {JHEP}\ }\textbf {\bibinfo {volume} {11}},\ \bibinfo {pages} {148} (\bibinfo
  {year} {2012})},\ \Eprint {http://arxiv.org/abs/1210.3340} {arXiv:1210.3340
  [hep-ph]} \BibitemShut {NoStop}%
%%CITATION = ARXIV:1210.3340;%%
\bibitem [{\citenamefont {DeWitt}(2003)}]{dewitt2003global}%
  \BibitemOpen
  \bibfield  {author} {\bibinfo {author} {\bibfnamefont {B.S.}\ \bibnamefont
  {DeWitt}},\ }\href@noop {} {\emph {\bibinfo {title} {The Global Approach to
  Quantum Field Theory Vol. 1}}},\ \bibinfo {series} {International series of
  monographs on physics}, Vol.\ \bibinfo {volume} {114}\ (\bibinfo  {publisher}
  {Clarendon Press},\ \bibinfo {year} {2003})\BibitemShut {NoStop}%
\bibitem [{\citenamefont {Buzzegoli}\ and\ \citenamefont
  {Becattini}(2018)}]{Buzzegoli:2018wpy}%
  \BibitemOpen
  \bibfield  {author} {\bibinfo {author} {\bibfnamefont {M.}~\bibnamefont
  {Buzzegoli}}\ and\ \bibinfo {author} {\bibfnamefont {F.}~\bibnamefont
  {Becattini}},\ }\bibfield  {title} {\enquote {\bibinfo {title} {{General
  thermodynamic equilibrium with axial chemical potential for the free Dirac
  field}},}\ }\href {\doibase 10.1007/JHEP12(2018)002} {\bibfield  {journal}
  {\bibinfo  {journal} {JHEP}\ }\textbf {\bibinfo {volume} {12}},\ \bibinfo
  {pages} {002} (\bibinfo {year} {2018})},\ \Eprint
  {http://arxiv.org/abs/1807.02071} {arXiv:1807.02071 [hep-th]} \BibitemShut
  {NoStop}%
%%CITATION = ARXIV:1807.02071;%%
\bibitem [{\citenamefont {Duff}(1994)}]{Duff:1993wm}%
  \BibitemOpen
  \bibfield  {author} {\bibinfo {author} {\bibfnamefont {M.~J.}\ \bibnamefont
  {Duff}},\ }\bibfield  {title} {\enquote {\bibinfo {title} {{Twenty years of
  the Weyl anomaly}},}\ }\bibfield  {booktitle} {\emph {\bibinfo {booktitle}
  {{Conference on Highlights of Particle and Condensed Matter Physics
  (SALAMFEST) Trieste, Italy, March 8-12, 1993}}},\ }\href {\doibase
  10.1088/0264-9381/11/6/004} {\bibfield  {journal} {\bibinfo  {journal}
  {Class. Quant. Grav.}\ }\textbf {\bibinfo {volume} {11}},\ \bibinfo {pages}
  {1387--1404} (\bibinfo {year} {1994})},\ \Eprint
  {http://arxiv.org/abs/hep-th/9308075} {arXiv:hep-th/9308075 [hep-th]}
  \BibitemShut {NoStop}%
%%CITATION = HEP-TH/9308075;%%
\bibitem [{\citenamefont {Eling}\ \emph {et~al.}(2013)\citenamefont {Eling},
  \citenamefont {Oz}, \citenamefont {Theisen},\ and\ \citenamefont
  {Yankielowicz}}]{Eling:2013bj}%
  \BibitemOpen
  \bibfield  {author} {\bibinfo {author} {\bibfnamefont {Christopher}\
  \bibnamefont {Eling}}, \bibinfo {author} {\bibfnamefont {Yaron}\ \bibnamefont
  {Oz}}, \bibinfo {author} {\bibfnamefont {Stefan}\ \bibnamefont {Theisen}}, \
  and\ \bibinfo {author} {\bibfnamefont {Shimon}\ \bibnamefont
  {Yankielowicz}},\ }\bibfield  {title} {\enquote {\bibinfo {title} {{Conformal
  Anomalies in Hydrodynamics}},}\ }\href {\doibase 10.1007/JHEP05(2013)037}
  {\bibfield  {journal} {\bibinfo  {journal} {JHEP}\ }\textbf {\bibinfo
  {volume} {05}},\ \bibinfo {pages} {037} (\bibinfo {year} {2013})},\ \Eprint
  {http://arxiv.org/abs/1301.3170} {arXiv:1301.3170 [hep-th]} \BibitemShut
  {NoStop}%
%%CITATION = ARXIV:1301.3170;%%
\bibitem [{\citenamefont {Fuini}\ and\ \citenamefont
  {Yaffe}(2015)}]{Fuini:2015hba}%
  \BibitemOpen
  \bibfield  {author} {\bibinfo {author} {\bibfnamefont {John~F.}\ \bibnamefont
  {Fuini}}\ and\ \bibinfo {author} {\bibfnamefont {Laurence~G.}\ \bibnamefont
  {Yaffe}},\ }\bibfield  {title} {\enquote {\bibinfo {title}
  {{Far-from-equilibrium dynamics of a strongly coupled non-Abelian plasma with
  non-zero charge density or external magnetic field}},}\ }\href {\doibase
  10.1007/JHEP07(2015)116} {\bibfield  {journal} {\bibinfo  {journal} {JHEP}\
  }\textbf {\bibinfo {volume} {07}},\ \bibinfo {pages} {116} (\bibinfo {year}
  {2015})},\ \Eprint {http://arxiv.org/abs/1503.07148} {arXiv:1503.07148
  [hep-th]} \BibitemShut {NoStop}%
%%CITATION = ARXIV:1503.07148;%%
\bibitem [{\citenamefont {Megias}\ and\ \citenamefont
  {Valle}(2014)}]{Megias:2014mba}%
  \BibitemOpen
  \bibfield  {author} {\bibinfo {author} {\bibfnamefont {Eugenio}\ \bibnamefont
  {Megias}}\ and\ \bibinfo {author} {\bibfnamefont {Manuel}\ \bibnamefont
  {Valle}},\ }\bibfield  {title} {\enquote {\bibinfo {title} {{Second-order
  partition function of a non-interacting chiral fluid in 3+1 dimensions}},}\
  }\href {\doibase 10.1007/JHEP11(2014)005} {\bibfield  {journal} {\bibinfo
  {journal} {JHEP}\ }\textbf {\bibinfo {volume} {11}},\ \bibinfo {pages} {005}
  (\bibinfo {year} {2014})},\ \Eprint {http://arxiv.org/abs/1408.0165}
  {arXiv:1408.0165 [hep-th]} \BibitemShut {NoStop}%
%%CITATION = ARXIV:1408.0165;%%
\bibitem [{\citenamefont {Buzzegoli}\ \emph {et~al.}(2017)\citenamefont
  {Buzzegoli}, \citenamefont {Grossi},\ and\ \citenamefont
  {Becattini}}]{Buzzegoli:2017cqy}%
  \BibitemOpen
  \bibfield  {author} {\bibinfo {author} {\bibfnamefont {M.}~\bibnamefont
  {Buzzegoli}}, \bibinfo {author} {\bibfnamefont {E.}~\bibnamefont {Grossi}}, \
  and\ \bibinfo {author} {\bibfnamefont {F.}~\bibnamefont {Becattini}},\
  }\bibfield  {title} {\enquote {\bibinfo {title} {{General equilibrium
  second-order hydrodynamic coefficients for free quantum fields}},}\ }\href
  {\doibase 10.1007/JHEP07(2018)119, 10.1007/JHEP10(2017)091} {\bibfield
  {journal} {\bibinfo  {journal} {JHEP}\ }\textbf {\bibinfo {volume} {10}},\
  \bibinfo {pages} {091} (\bibinfo {year} {2017})},\ \bibinfo {note} {[Erratum:
  JHEP07,119(2018)]},\ \Eprint {http://arxiv.org/abs/1704.02808}
  {arXiv:1704.02808 [hep-th]} \BibitemShut {NoStop}%
%%CITATION = ARXIV:1704.02808;%%
\bibitem [{\citenamefont {Kovtun}\ and\ \citenamefont
  {Shukla}(2019)}]{Kovtun:2019wjz}%
  \BibitemOpen
  \bibfield  {author} {\bibinfo {author} {\bibfnamefont {Pavel}\ \bibnamefont
  {Kovtun}}\ and\ \bibinfo {author} {\bibfnamefont {Ashish}\ \bibnamefont
  {Shukla}},\ }\bibfield  {title} {\enquote {\bibinfo {title} {{Einstein's
  Equations in Matter}},}\ }\href@noop {} {\  (\bibinfo {year} {2019})},\
  \Eprint {http://arxiv.org/abs/1907.04976} {arXiv:1907.04976 [gr-qc]}
  \BibitemShut {NoStop}%
%%CITATION = ARXIV:1907.04976;%%
\bibitem [{\citenamefont {Kapusta}\ and\ \citenamefont
  {Gale}(2011)}]{Kapusta:2006pm}%
  \BibitemOpen
  \bibfield  {author} {\bibinfo {author} {\bibfnamefont {J.~I.}\ \bibnamefont
  {Kapusta}}\ and\ \bibinfo {author} {\bibfnamefont {Charles}\ \bibnamefont
  {Gale}},\ }\href {\doibase 10.1017/CBO9780511535130} {\emph {\bibinfo {title}
  {{Finite-temperature field theory: Principles and applications}}}},\
  Cambridge Monographs on Mathematical Physics\ (\bibinfo  {publisher}
  {Cambridge University Press},\ \bibinfo {year} {2011})\BibitemShut {NoStop}%
%%CITATION = INSPIRE-738588;%%
\bibitem [{\citenamefont {Bellac}(2011)}]{Bellac:2011kqa}%
  \BibitemOpen
  \bibfield  {author} {\bibinfo {author} {\bibfnamefont {Michel~Le}\
  \bibnamefont {Bellac}},\ }\href {\doibase 10.1017/CBO9780511721700} {\emph
  {\bibinfo {title} {{Thermal Field Theory}}}},\ Cambridge Monographs on
  Mathematical Physics\ (\bibinfo  {publisher} {Cambridge University Press},\
  \bibinfo {year} {2011})\BibitemShut {NoStop}%
%%CITATION = INSPIRE-1384874;%%
\end{thebibliography}%
\onecolumngrid

\end{document}